\documentclass[usenatbib,a4paper,times,fleqn]{mnras}

\usepackage{graphicx,natbib,times,amssymb,amsmath,pstricks,soul,color}
\usepackage{aas_macros}
\usepackage{bm,url}
\usepackage{textcomp}

\usepackage{subfig}
\usepackage{mathtools}

\newcommand  *{\B}[1] {\boldsymbol{#1}}
\renewcommand*{\H}[1] {\hat{\B{#1}}}

\title[On The Secular Evolution of GG Tau A Circumbinary Disc]{On The Secular Evolution of GG Tau A Circumbinary Disc: A Msialigned Disc Scenario}
\author[Hossam Aly, Giuseppe Lodato, \& Paolo Cazzoletti]
       {Hossam Aly$^1$\thanks{Email: hossam.aly@unimi.it}
        Giuseppe Lodato$^1$,  and Paolo Cazzoletti$^2$\vspace{0.05in}\\ 
        $^1$ Universitá degli Studi di Milano, Via Giovanni Celoria 16, 20133 Milano, Italy\\
        $^2$ Max-Planck-Institut f{\"u}r Extraterrestrische Physik, Gie\ss enbachstra\ss e, 85741 Garching bei M{\"u}nchen, Germany}
\date{Accepted .
      Received ;
      }
\pagerange{\pageref{firstpage}--\pageref{lastpage}}
\pubyear{2015}
\begin{document}
\maketitle
\label{firstpage}
%
%
\begin{abstract}
The binary system GG Tau A is observed to have a circumbinary disc with a dust ring located further out than expected, assuming a co-planar disc and a corresponding semi-major axis of 34 AU. Given the binary separation, this large cavity can be explained by relaxing the assumption of a co-planar disc and instead fit the observations with a mis-aligned circumbinary disc around an eccentric binary with a wider semi-major axis of 60 AU, consistent with fitting the proper motion data for the system. We run SPH simulations to check this possibility and indeed we find that a misalignment angle of 30 degrees and a binary eccentricity of 0.45 fit both the astrometric data and the disc cavity. However, such configuration could in principle be unstable to polar, rather that planar alignment. We investigate the secular evolution of this configuration and show that it is indeed stable throughout the disc lifetime. 

\end{abstract}

\begin{keywords}
accretion, accretion discs -- hydrodynamics -- stars: formation -- stars: individual: GG Tau A -- binaries: close

\end{keywords}

\section{Introduction}
Young stellar objects are often observed to be accompanied by circumstellar discs, which form as a result of the conservation of angular momentum of the collapsing cloud core. These discs of gas and dust are thought to be the cradle in which planets form. Since many stars form in binaries, we expect many of them to have circumbinary discs as well as circumbinary planets. Indeed, a few planets (11 to date) have already been observed around binary systems \citep{DoyleEtal2011,WelshEtal2012,OroszEtal2012a,OroszEtal2012b,WelshEtal2015,KostovEtal2013,KostovEtal2016}. Circumbinary discs are subject to binary-disc interactions that significantly affect their evolution \citep{Papaloizoun&Pringle1977}; in particular, while disc viscosity drives mass inwards, tidal torques rip material from the inner parts of the disc. These two effects are balanced at the tidal truncation radius and a cavity is opened \citep{Lin&Papaloizou1986}. The extent of the cavity and its dependence on the disc viscosity and binary orbital parameters has been theoretically estimated by \citet{Artymowicz&Lubow1994} for co-planar discs. \citet{Miranda&Lai2015} and \citet{LubowEtal2015} worked out the effect of tidal torques due to Lindblad resonances for misaligned discs and found it to be much weaker than in the case of co-planar discs. Thus we expect misaligned circumbinary discs to extend further in than co-planar ones. Retrograde discs do not experience such resonances, but in the case of an eccentric binary there exist resonances due to the retrogradely rotating components of the binary potential \citep{Nixon&Lubow2015}.

The quadruple stellar system GG Tau, for which recent observations are readily available, provides a prime example where these processes take place (a possible 5th component has been reported by \citet{DiFolcoEtal2014}). This system comprises 2 binary systems; GG Tau A and GG Tau B, with the former observed to have a circumbinary disc. Due to its relative proximity ($\sim$ 140 pc), this system has been widely studied and proper motions for the stellar components are available. \citet{AndrewsEtal2014} modeled observations of continuum dust emission at 4 different mm wavelengths and concluded that the binary is encircled by a circumbinary dust ring at an average radius of 235 AU with a FWHM of 60 AU. This finding is also consistent with more recent ALMA 0.9 mm observations by \citet{TangEtal2016} 

\citet{CazzolettiEtal2017} investigated how can the recent dust ring observation of \citet{AndrewsEtal2014} be explained in light of theoretical predictions provided by tidal truncation theory. Fixing one orbital parameter at a time, one can constrain the others by fitting proper motion data available for the system \citep{Khoeler2011}. \citet{CazzolettiEtal2017} focused on the case of a co-planar disc, for which the binary is constrained to have a semi major axis $a=34$ AU and an eccentricity $e=0.28$. The expected tidal truncation radius for such a configuration is expected to be $\sim$ 2 to 3 times the semi major axis. This is seemingly in conflict with \citet{AndrewsEtal2014} dust ring observation which has a radius far exceeding the predicted inner disc radius. \citet{CazzolettiEtal2017} performed hydrodynamical simulations of the co-planar case and indeed found that the pressure bump in the disc, in which the dust ring is expected to be centered, is located at radii less than 150 AU. They conclude that the GG Tau A dust ring is not consistent with a co-planar disc configuration.

Relaxing the assumption of co-planarity, one can investigate the possibility of a wider binary with a misaligned disc. \citet{Khoeler2011} shows that for a semi major axis $a=60$ AU, the astrometric data is consistent with a disc inclination of $\sim$ 25 degrees and a binary eccentricity of 0.44. At a first glance this configuration seems more promising to alleviate the conflict with the dust ring observation. However, the picture for a misaligned circumbinary disc is further complicated by the presence of binary torques causing differential precession in the disc around the binary angular momentum vector, leading to the evolution of warps in the disc \citep{Papaloizou&Pringle1983,Papaloizou&Lin1995,Ogilivie1999,Lodato&Pringle2007,Nixon&King2016,DoganEtal2018}. Depending on the disc viscosity, this induced warp either diffuses or propagates in the disc in a wave-like manner. The final configuration for a viscous circumbinary disc around a \emph{circular} binary is either a co- or counter-aligned disc with respect to the binary, depending on the initial misalignment angle and the disc angular momentum compared to that of the binary \citep{NixonEtal2011,Nixon&King2012,KingEtal2005}.

\begin{figure}
\resizebox{80.0mm}{!}{\mbox{\includegraphics[angle=-90]{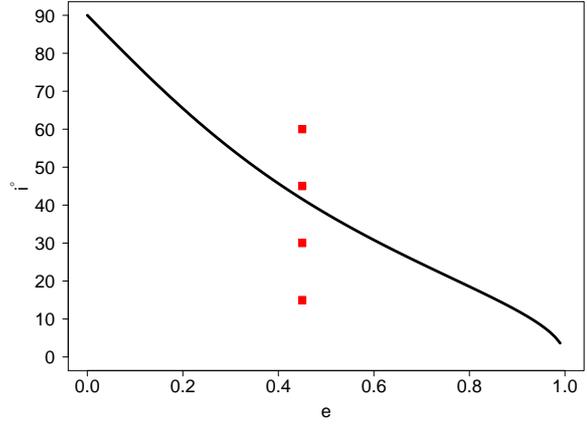}}}
\caption{\label{fig:critical}
    Critical inclination above which polar alignment occurs, as a function of binary eccentricities, assuming a negligible disc mass. The points are the 4 different initial disc inclinations considered in this paper}
\end{figure}

For the more general case of a misaligned disc around an eccentric binary, \citet{AlyEtal2015} showed analytically that disc-binary interaction causes the disc to differentially precess around both the pole of the binary plane and the binary eccentricity vector. The latter component of the precession vector may lead to polar alignment around the binary eccentricity vector for highly eccentric binaries, depending on the initial misalignment angle. In particular, they find that considering the interaction between a circumbinary circular ring and eccentric binary up to quadrupole order, the precession will be azimuthal; i.e, around the binary angular momentum vector, if this condition is met:
\begin{equation} \label{eq:condition}
	(1-e^2)(\H{l}\cdot\H{h})^2 > 5e^2(\H{l}\cdot\H{e})^2
\end{equation}
where $\H{l}$ is the ring specific angular momentum unit vector, $\B{h}$ is the binary specific angular momentum vector, and $\B{e}$ is the binary eccentricity vector.
If this condition is not met then polar precession around the binary eccentricity vector will take place instead. Choosing a configuration in which the the ring line of nodes is perpendicular to the binary eccentricity vector, this condition can be re-written in the form of a critical inclination angle, above which polar precession occurs (also displayed in Fig.~\ref{fig:critical}):
\begin{equation} \label{eq:incl}
	i^\circ=\tan^{-1}\sqrt{\frac{1-e^2}{5e^2}}
\end{equation}
Moreover, for higher values of binary eccentricities, the disc will experience inclination oscillations until a final configuration is reached. These oscillations are related to the well known Kozai-Lidov mechanism for the restricted three-body problem when taking the massless body to be the outer one \citep{Verrier&Evans2009, Farago&Laskar2010, Doolin&Blundell2011}. Note that this analysis applies only to low mass discs so that the co-evolution of the binary can be neglected. For discs with comparable mass to the binary numerical simulations are required to determine their dynamical behaviour.

\citet{AlyEtal2015} performed hydrodynamical simulations which confirmed this evolution in the case of low mass thin discs around supermassive black hole binaries. \citet{Martin&Lubow2017} showed that this is also the case for low mass thick protostellar discs. Applying equation~(\ref{eq:incl}) to an eccentricity $e=0.45$; i.e, the eccentricity chosen for our misaligned case, we get a critical inclination $i\sim42^\circ$ above which a low mass disc is expected to polar-align. We note that this critical inclination is somewhat close to the range of inclinations obtained from orbital fitting.

In this paper, we study the evolution of a misaligned gas disc around GG Tau A aiming to resolve the apparent conflict with the dust ring observation. We perform Smoothed Particle Hydrodynamics (SPH) simulations, fixing the binary semi major axis to $a=60$ and eccentricity to $e=0.45$ (consistent with \citealt{Khoeler2011} orbital solutions), and varying the disc initial misalignment and mass. We present the details of the numerical simulations in Section~\ref{sec:setup}, show our results Section~\ref{sec:results}, and finally conclude our findings in Section~\ref{sec:conclusion}.

\begin{figure*}
  \begin{center}
    \resizebox{80.0mm}{!}{\mbox{\includegraphics[angle=-90]{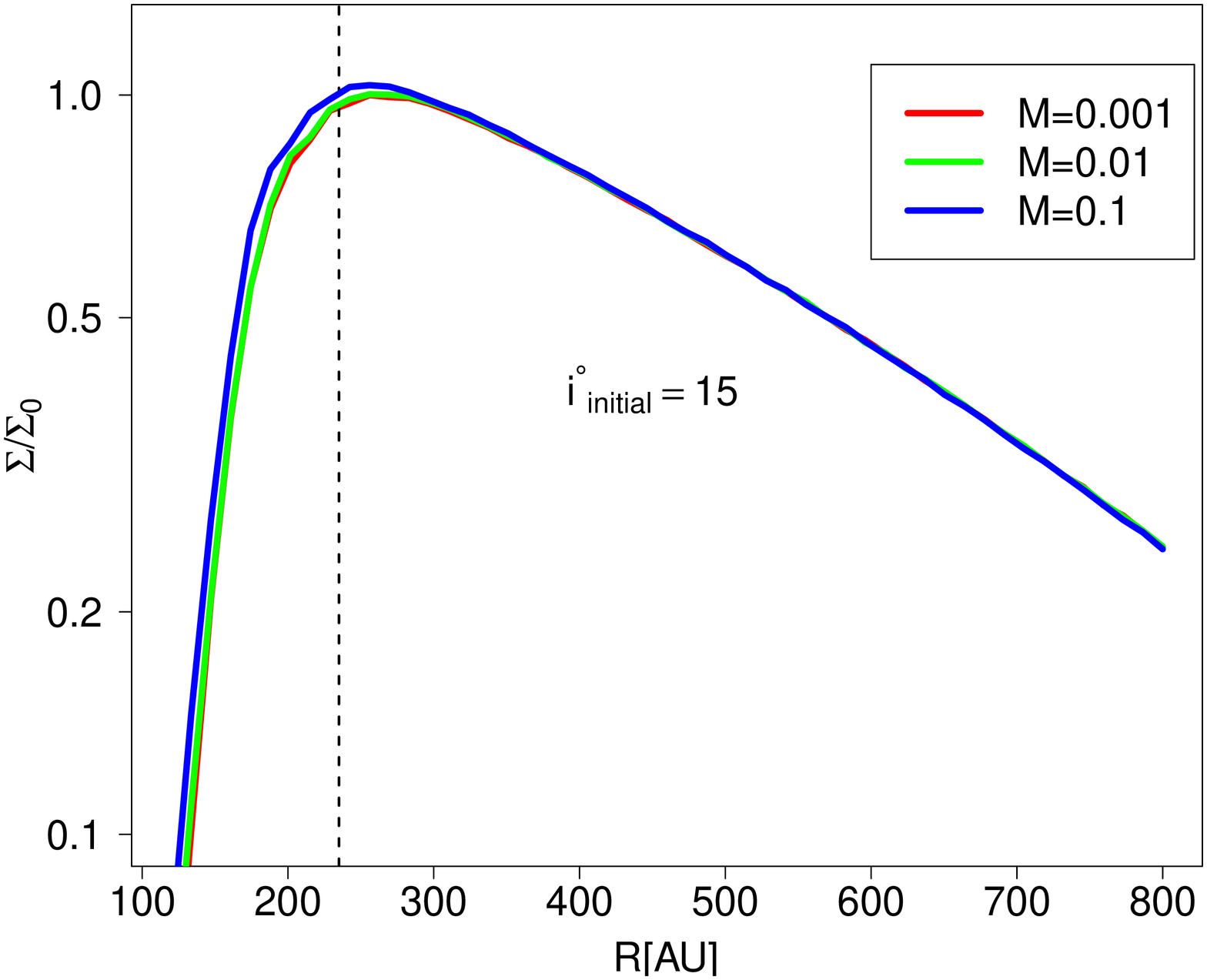}}} 
    \resizebox{80.0mm}{!}{\mbox{\includegraphics[angle=-90]{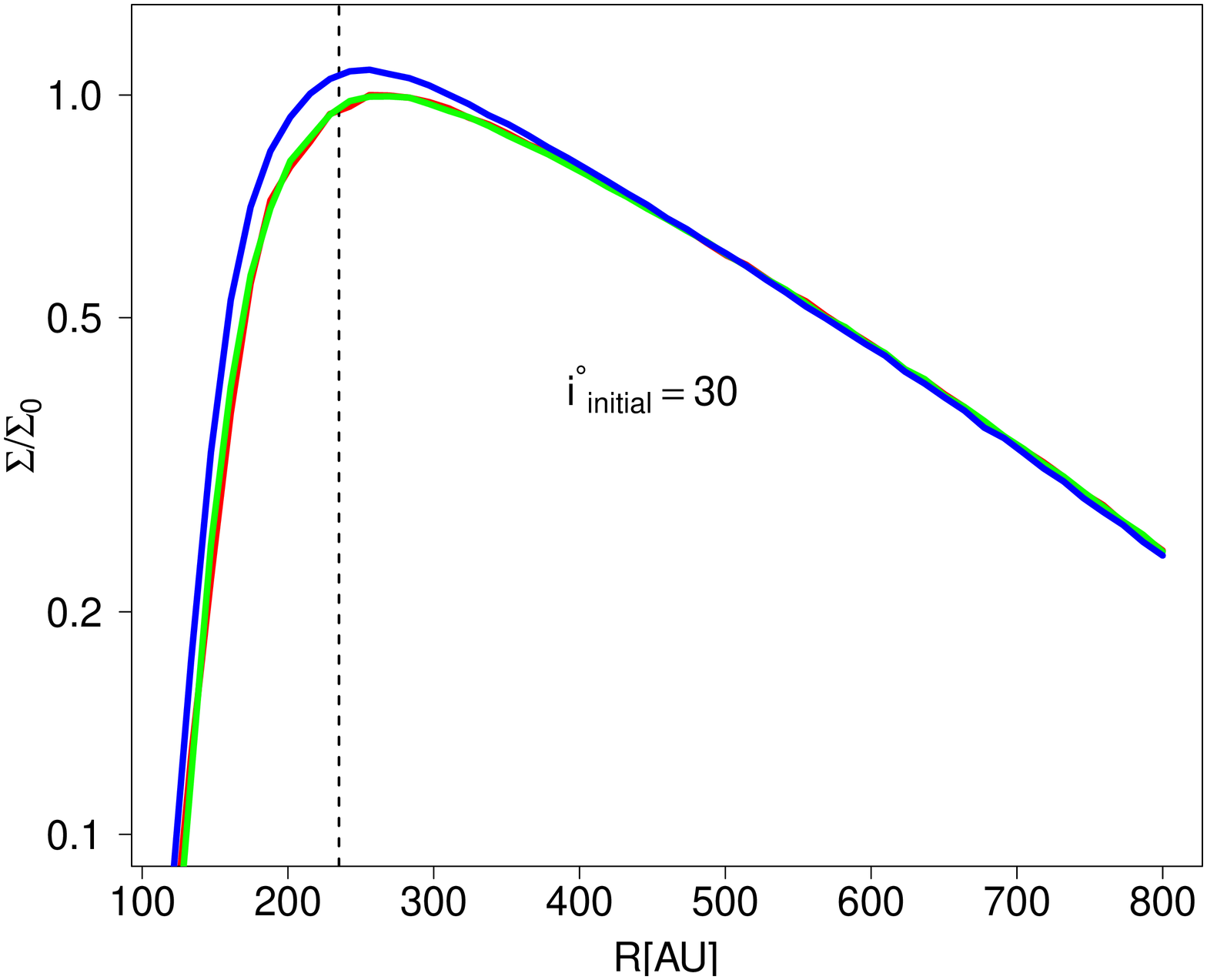}}} 
    \resizebox{80.0mm}{!}{\mbox{\includegraphics[angle=-90]{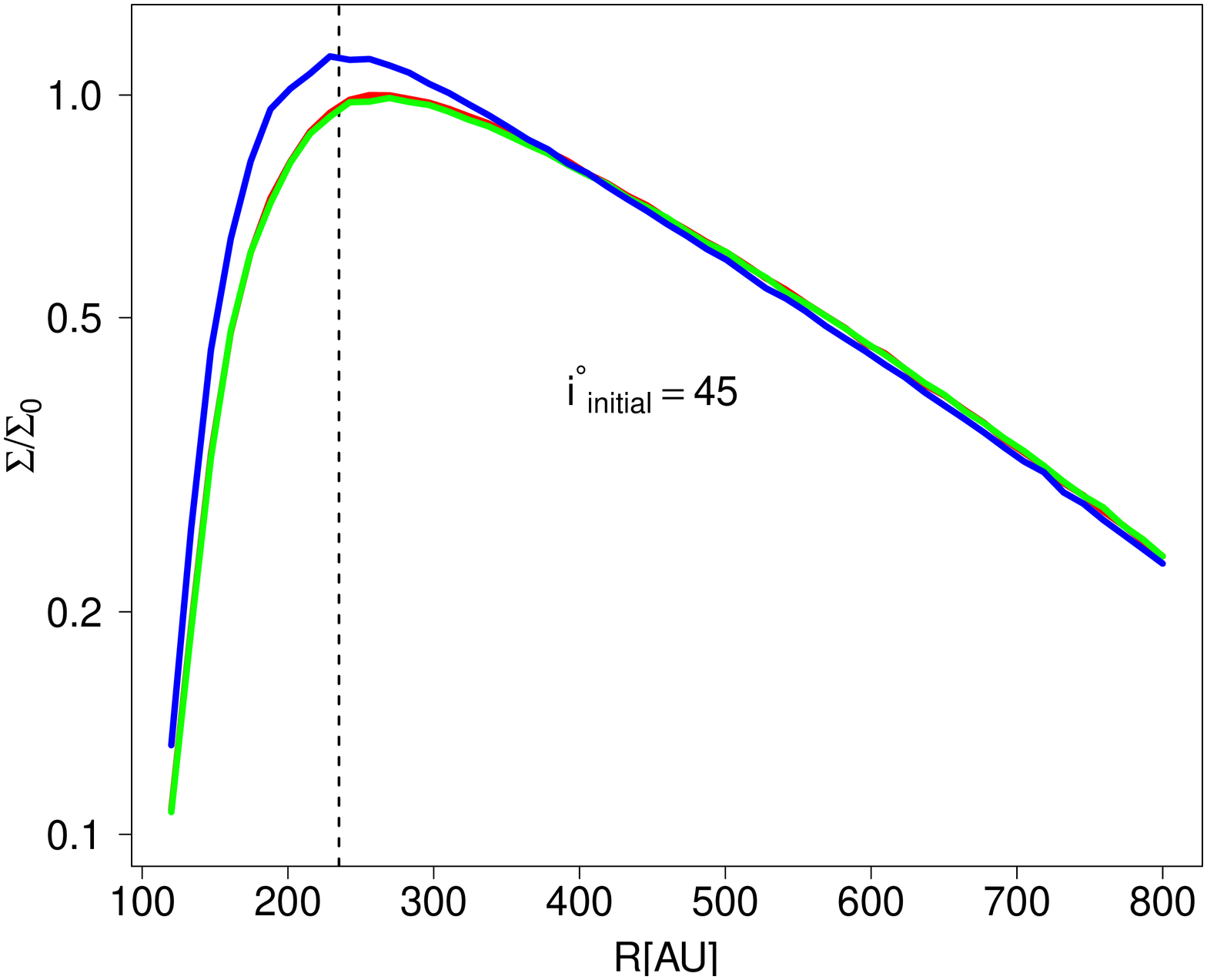}}}
    \resizebox{80.0mm}{!}{\mbox{\includegraphics[angle=-90]{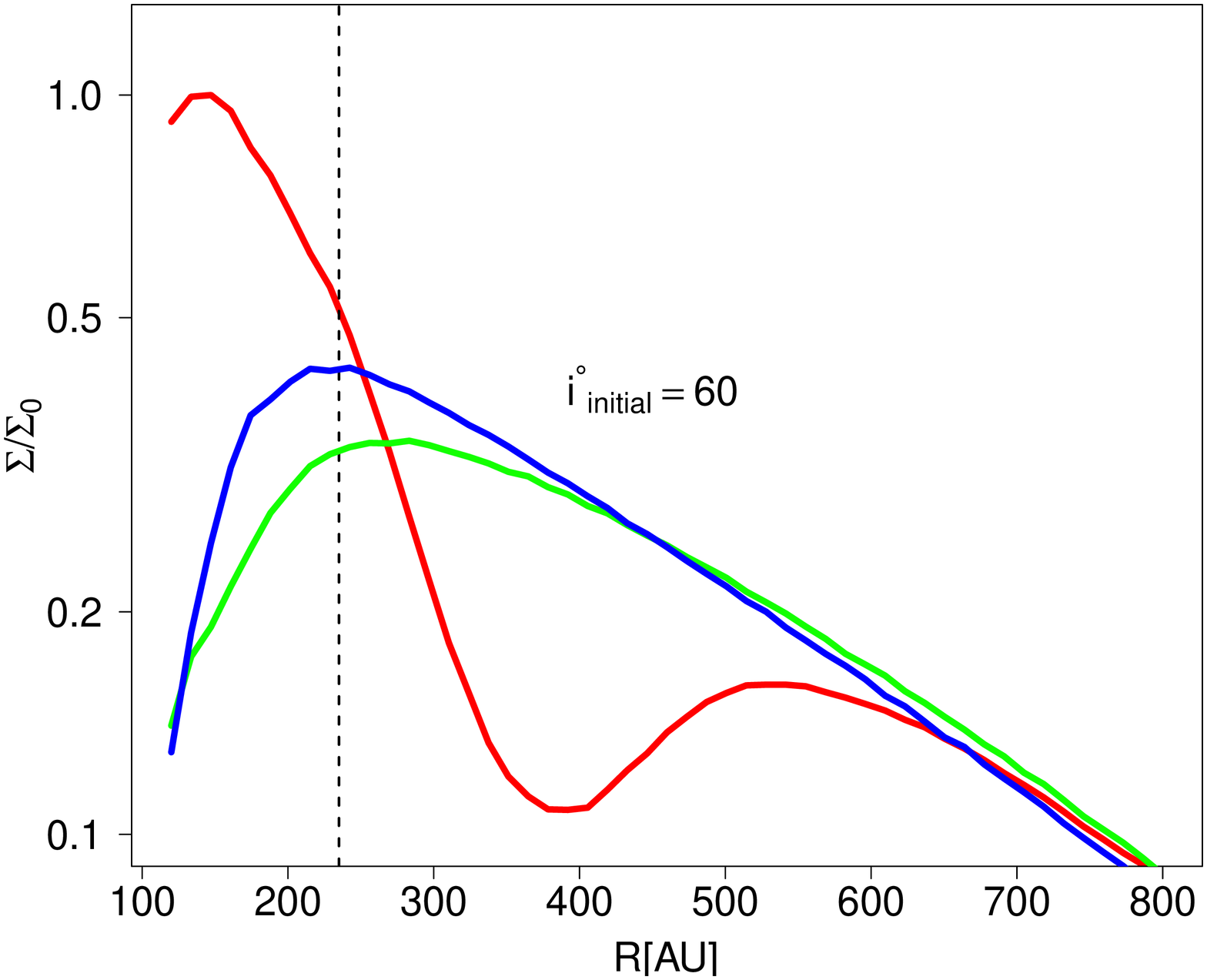}}}
 \caption{\label{fig:sigma}
    Radial surface density profiles at time t=1000 binary orbits for the entire parameter space. The binary parameters in all panels are $a=60$ AU and $e=0.45$. The dashed vertical line is the location of the dust ring inferred from \citet{AndrewsEtal2014} observations}
  \end{center}
\end{figure*}

\begin{figure*}
  \begin{center}
    \resizebox{80.0mm}{!}{\mbox{\includegraphics[angle=-90]{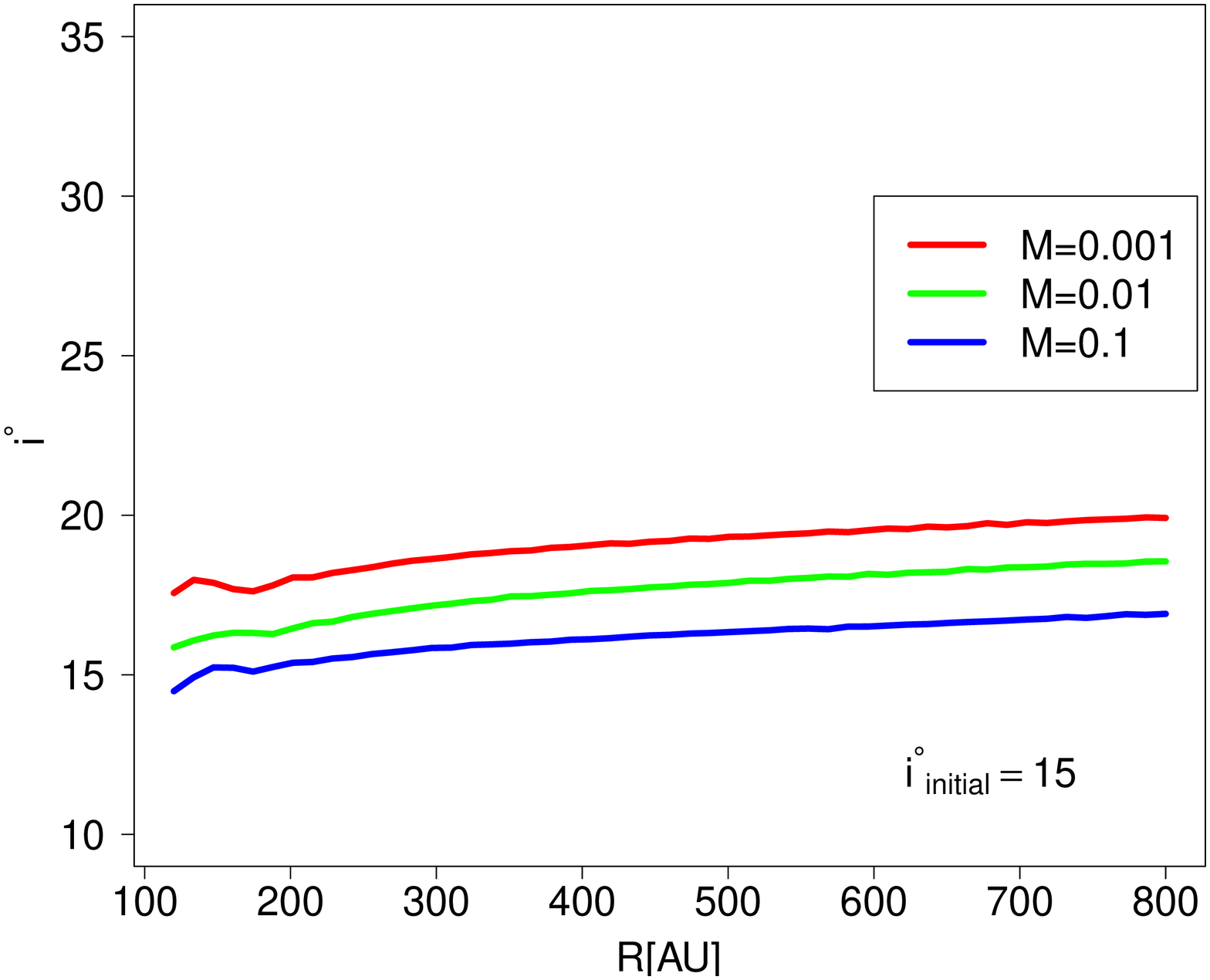}}} 
    \resizebox{80.0mm}{!}{\mbox{\includegraphics[angle=-90]{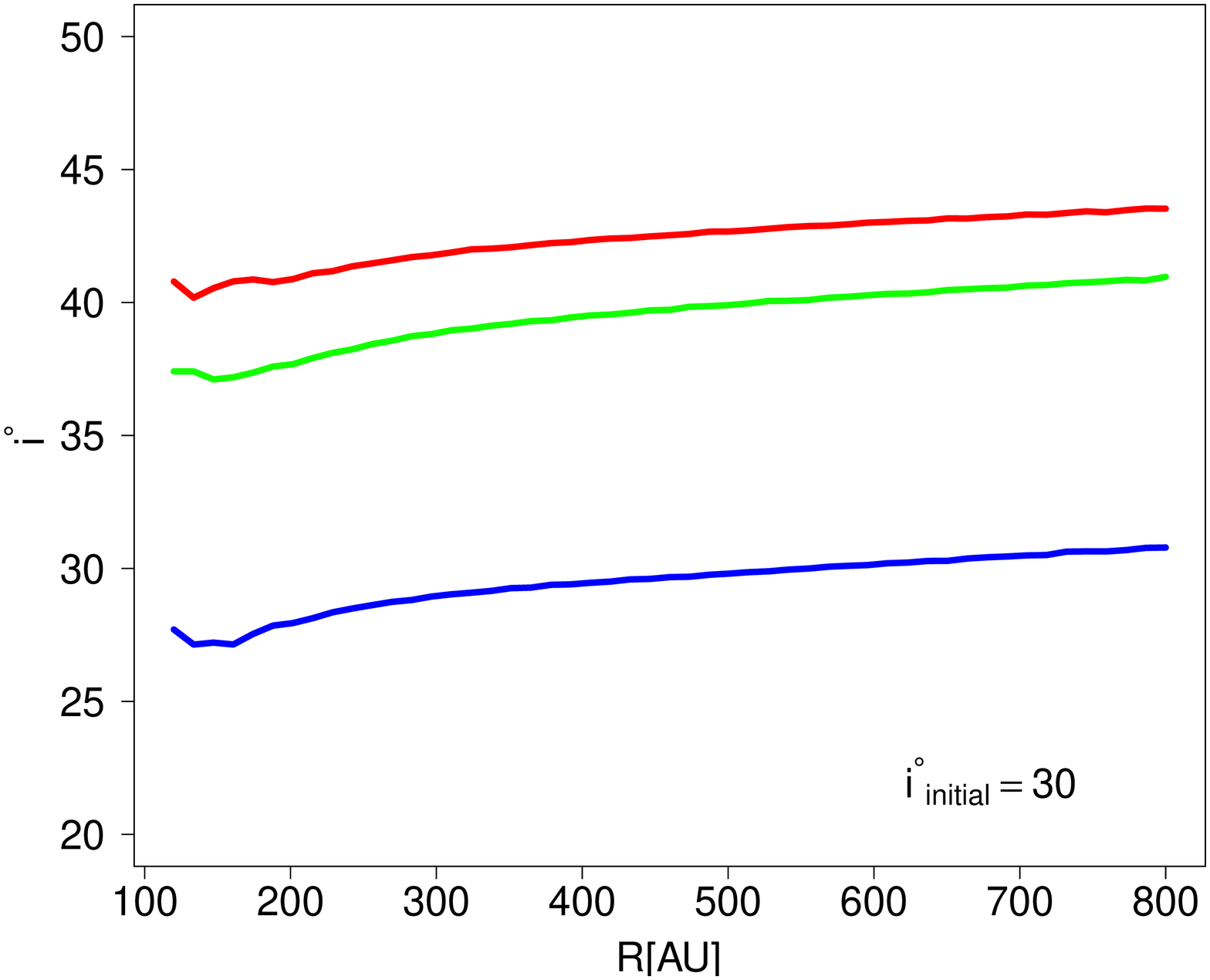}}} 
    \resizebox{80.0mm}{!}{\mbox{\includegraphics[angle=-90]{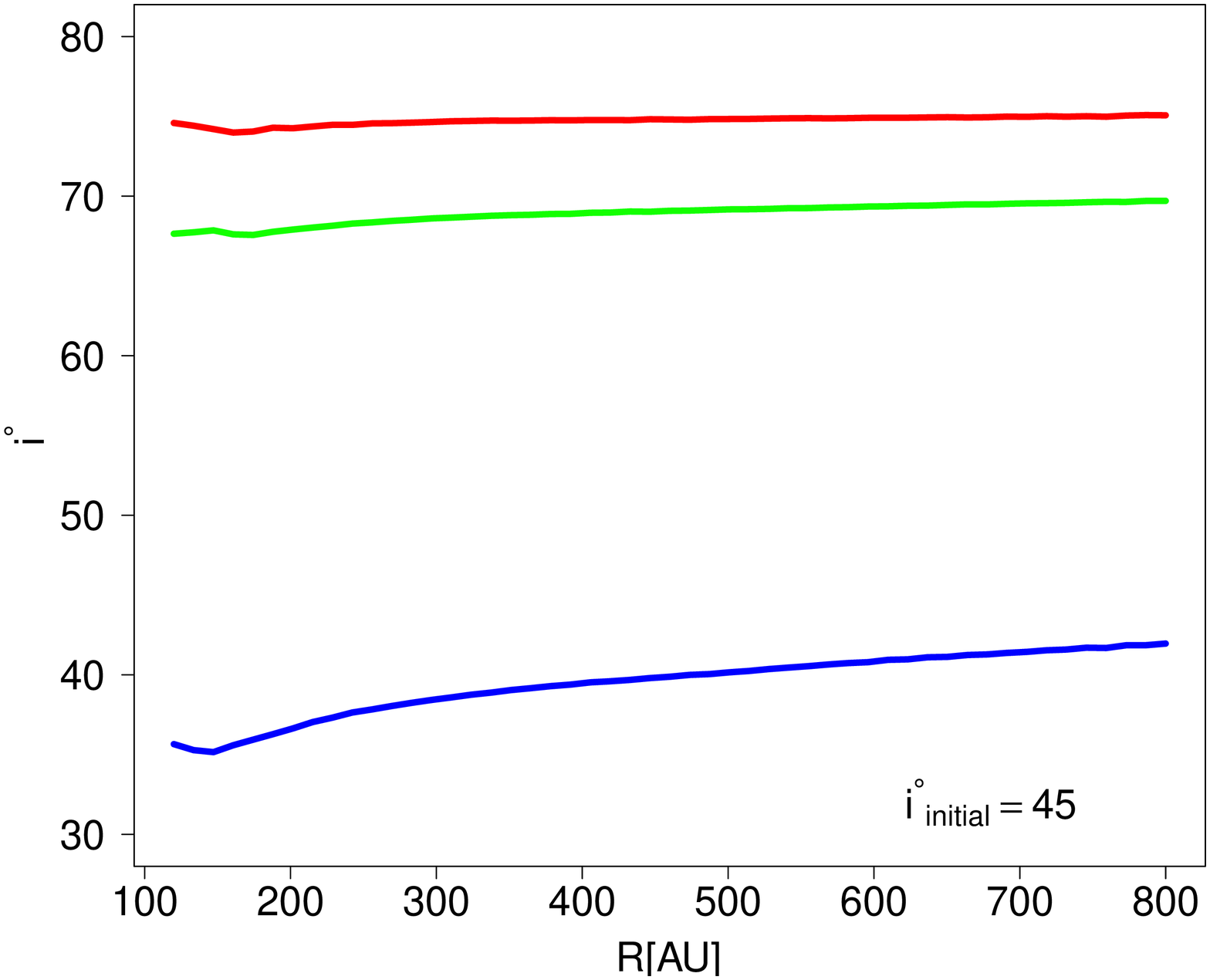}}}
    \resizebox{80.0mm}{!}{\mbox{\includegraphics[angle=-90]{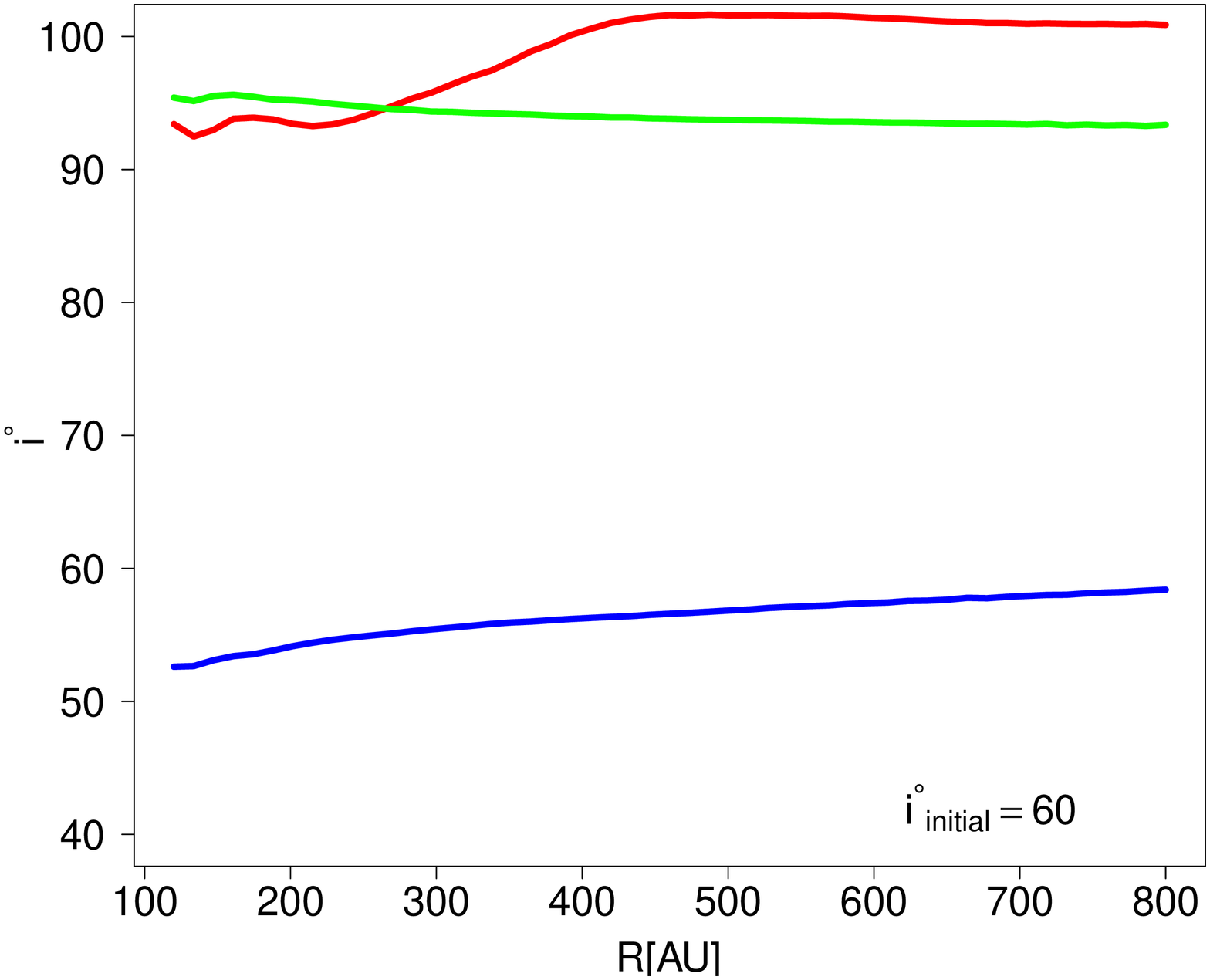}}}
 \caption{\label{fig:warp}
    Radial tilt profiles at time t=1000 binary orbits for the entire parameter space. The binary parameters are the same as Fig.~\ref{fig:sigma}}
  \end{center}
\end{figure*}

\begin{figure*}
  \begin{center}
    \resizebox{80.0mm}{!}{\mbox{\includegraphics[angle=0]{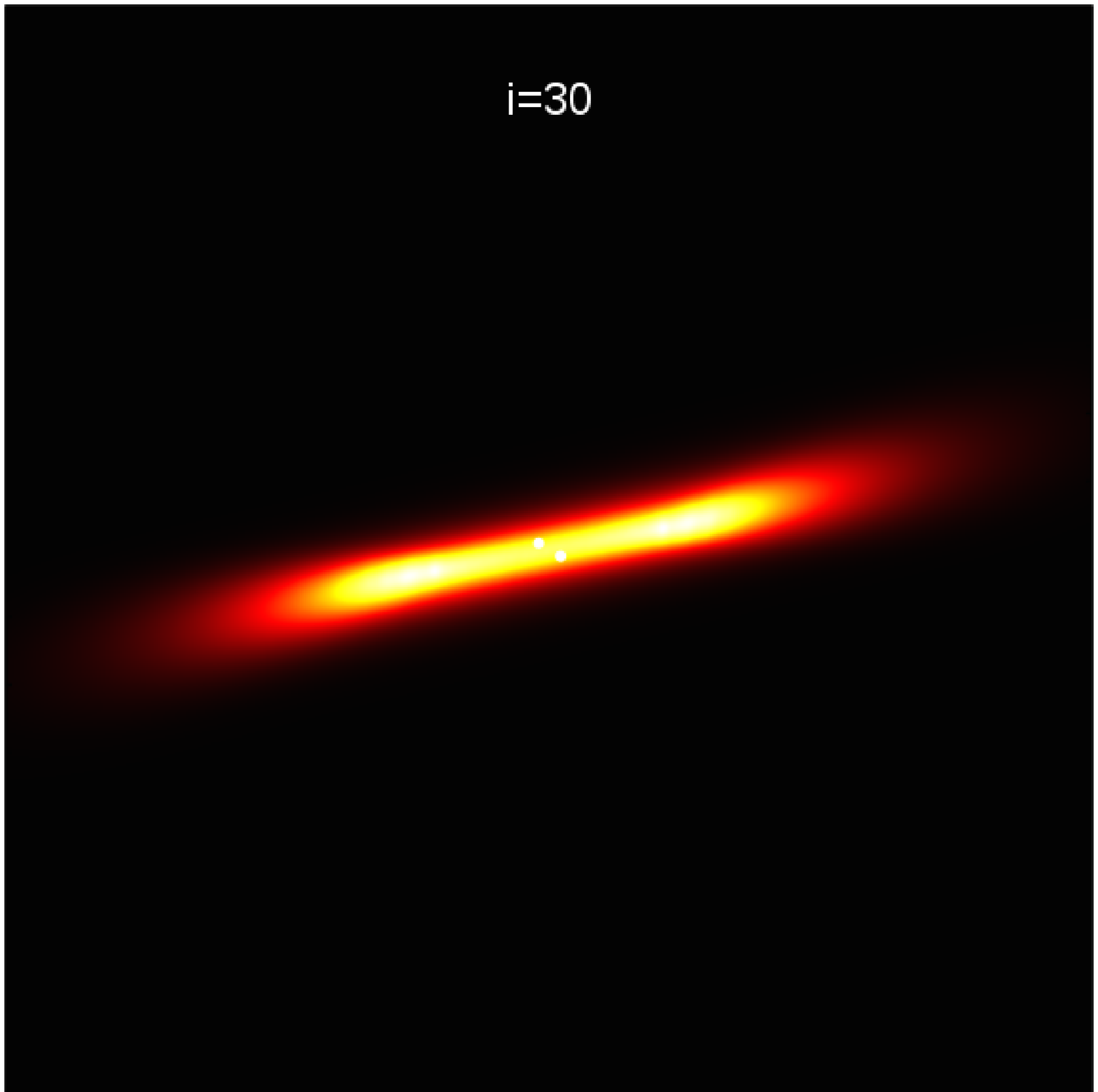}}} 
    \resizebox{80.0mm}{!}{\mbox{\includegraphics[angle=0]{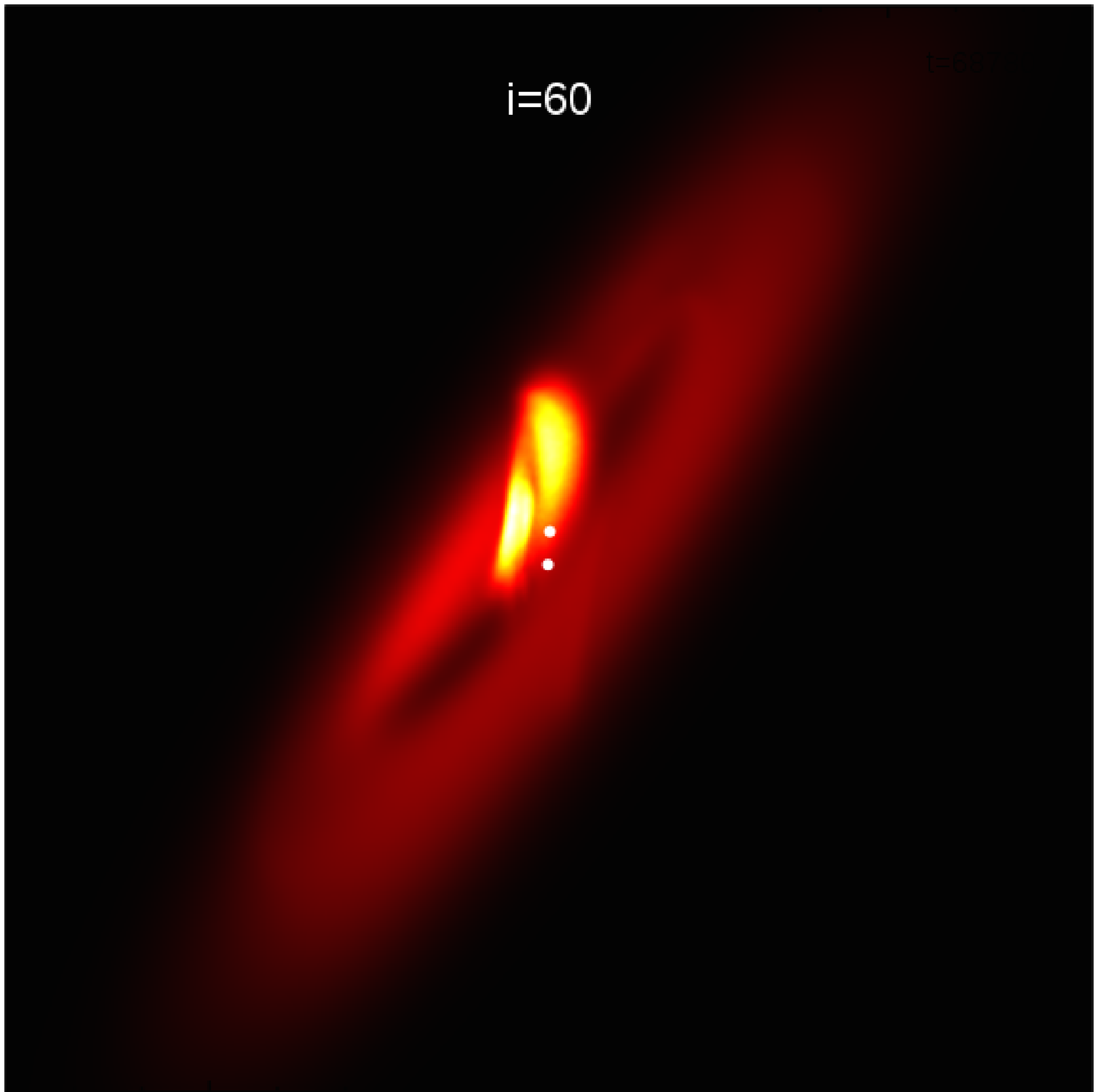}}} 
    \resizebox{80.0mm}{!}{\mbox{\includegraphics[angle=0]{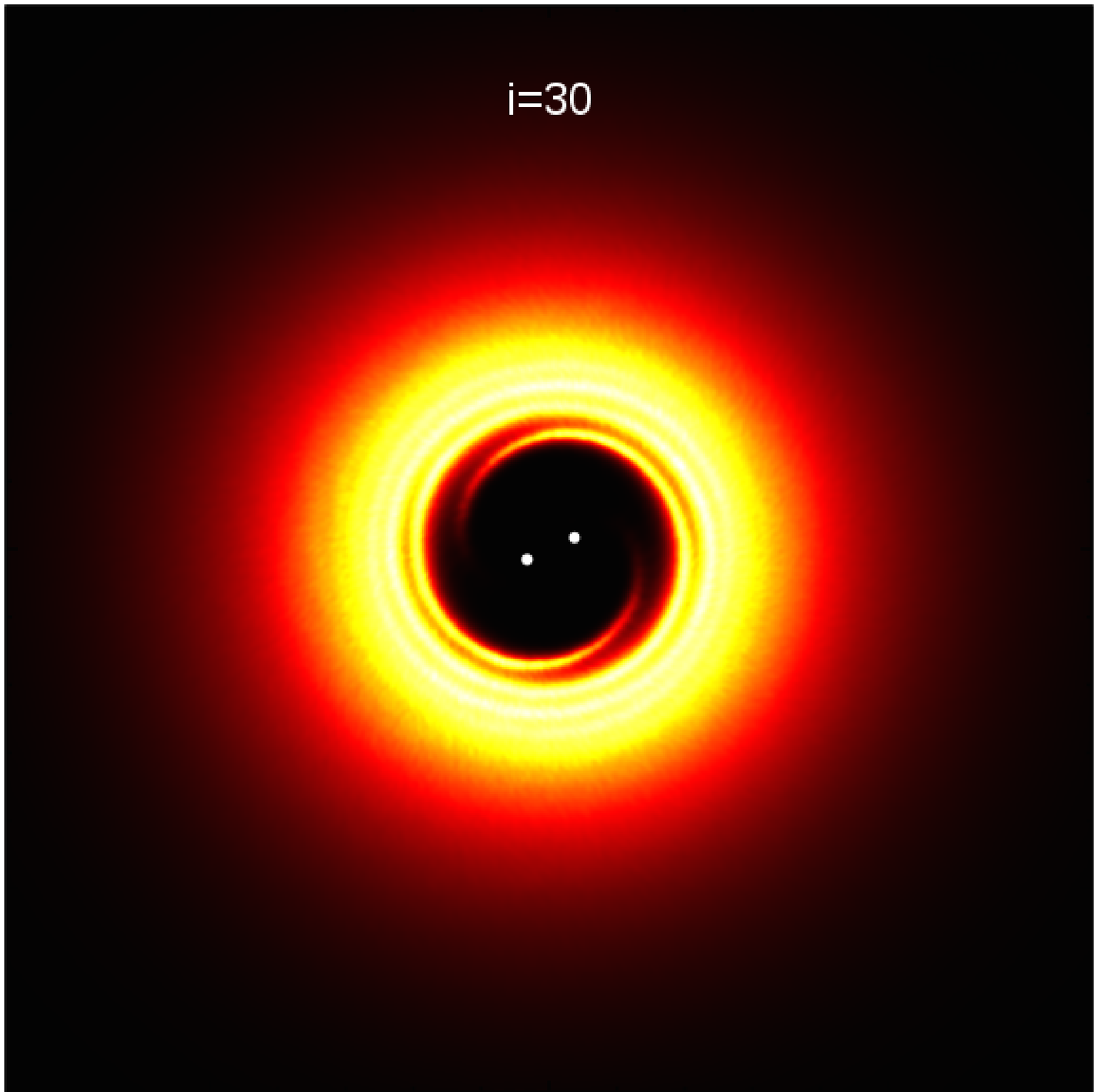}}}
    \resizebox{80.0mm}{!}{\mbox{\includegraphics[angle=0]{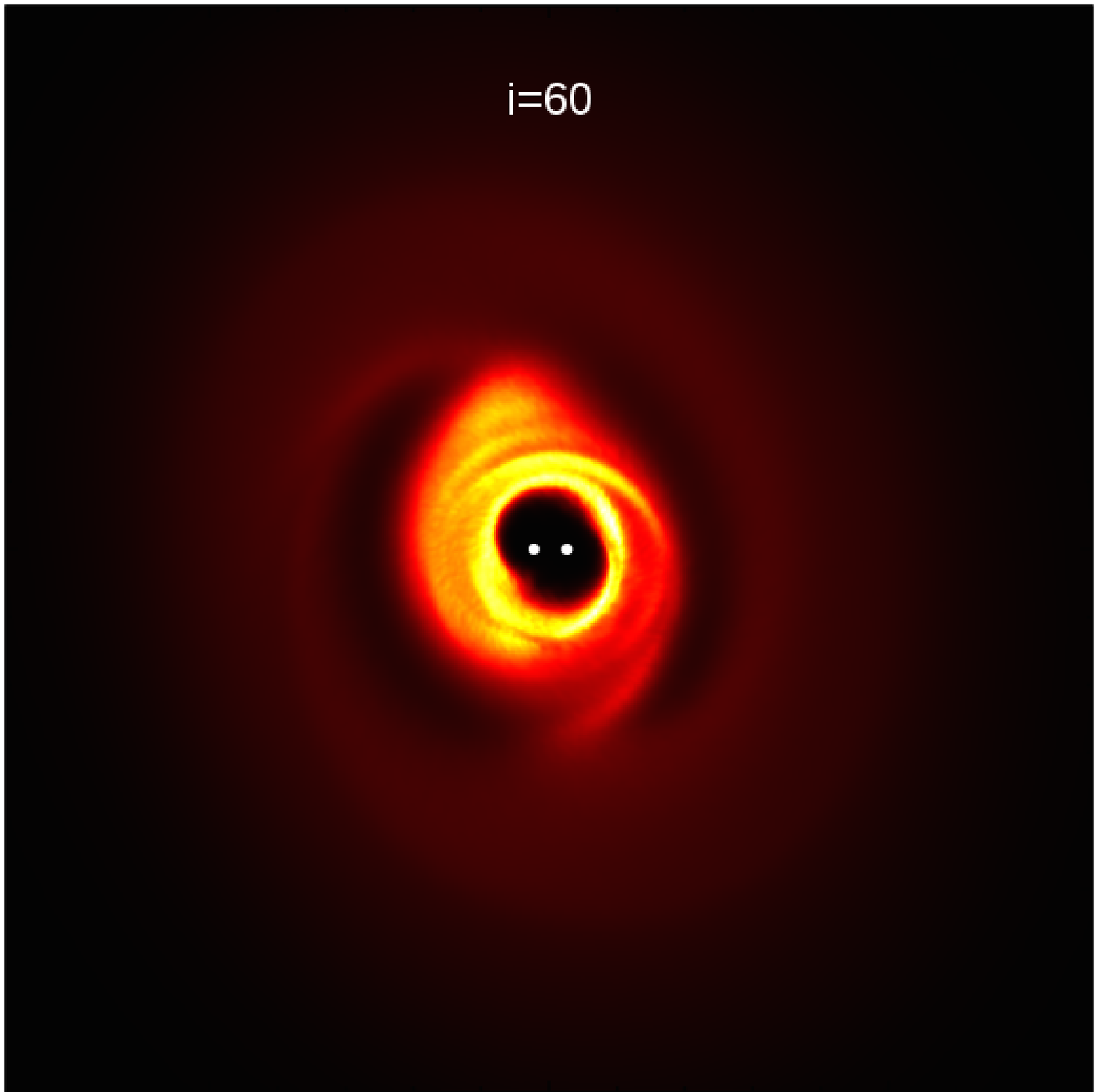}}}
      \end{center}
 \caption{\label{fig:splash}
    Projected density for 2 snapshots at time t=900 binary orbits for a disc with $M_d=0.001$ and inclinations $i=30^\circ$ and $60^\circ$} (left and right, respectively). Top (bottom) row is an edge (face) view.

\end{figure*}

\section{Simulations Setup}
We perform a suite of 3-D SPH simulations of misaligned circumbinary discs around the GG Tau A binary. We use the SPH PHANTOM \citep{PriceEtal2017} which has been used extensively for modeling accretion discs \citep[e.g.][]{LodatoPrice2010,NixonEtal2013,FacchiniEtal2013}. For all our inclined disc simulations we fix the binary semi major axis and eccentricity to $a=60$ and $e=0.45$. For comparison purposes, we run one simulation with a planar disc and a binary semi-major axis and eccentricity $a=34$ and $e=0.28$. We choose initial inclinations of ($i=15^\circ$, $30^\circ$, $45^\circ$, $60^\circ$) and disc masses of ($M_{\rm d}=0.1$, 0.01, 0.001) M$_\odot$. The disc initial inclinations are chosen so that we cover a wide range around $i=25^\circ$ obtained from fitting the proper motion data for the system with a wider semi-major axis. We run all simulations for 2000 binary orbits, which corresponds to $\sim 1.5\times10^5$ years. We use a disc viscosity coefficient \citep{ShakuraSunyaev1973} $\alpha=0.01$ implemented in the code with a physical viscosity treatment rather than mapping it to artificial viscosity \citep{LodatoPrice2010}. 

We model the binary stellar components using sink particles with masses of 0.78 and 0.68 M$_\odot$ \citep{WhiteEtal1999}. Both sink particles have an accretion radius of $r_{\rm acc}=0.2a$ inside which particles are removed from the simulation to avoid exceedingly short timesteps, since we are not interested in resolving circumprimary and circumsecondary discs around either components (accreted particles have their masses and momentum added to the corresponding sink particle). We model the disc with an initial inner and outer radii $R_{\rm in}=120$ and $R_{\rm out}=800$ AU (chosen as such to limit the computational time for each simulation; test runs with a larger outer radius of 1200 AU produced similar results). The disc aspect ratio $H/R$ is chosen to be 0.12 at the inner radius, which results in a gravitationally stable disc for all the mass range in our parametric study, and hence gas self gravity is not included. We use 1 million SPH particles in all our simulations, which for our choice of dynamical range ensures that the disc scale height is resolved by a few resolution lengths.

In all our simulations we use a gas density radial profile as a power law with an index $p=-1$. We assume a locally isothermal disc, where the sound speed $c_s$ varies only with disc radius as power law with an index of $-0.45$. This corresponds to a temperature profile of the form \citep{GuilloteauEtal1999,DutreyEtal2014,TangEtal2016}:
\begin{equation}
    T=20\mathrm{K}\bigg(\frac{R}{300\mathrm{AU}}\bigg)^{-0.9}
\end{equation}

\label{sec:setup}

\section{Results and Discussions}
\label{sec:results}
In Fig.~\ref{fig:sigma} we present the radial density profiles for the entire simulations suite (12 in total) at time t=1000 binary orbits. Each panel shows the results for a different initial inclination ($i=15^\circ$, $30^\circ$, $45^\circ$, $60^\circ$), while varying disc masses ($M_{\rm d}=0.001$, 0.01, 0.1) M$_\odot$. It is clear that in most cases the maximum density (and hence the pressure bump) is located at $\sim$ 235 AU, consistent with the observed dust ring \citep{PinillaEtal2012}. The only exception is the disc with $i=60^\circ$ inclination and low mass ($M_{\rm d}=0.001$). In this case, as the inner disc tends to reach a polar configuration, the warp magnitude becomes significant and the disc breaks into various rings (also clear from the column density snapshots in Fig.~\ref{fig:splash}, as well as Fig.~\ref{fig:warp} and Fig.~\ref{fig:tilt}, see below), as observed in \citet{AlyEtal2015} simulations for similar parameters. For the medium mass case, a disc break is not evident, but as the inner part becomes polar, and hence the cavity shrinks, the density peak moves inwards. We note that apart from these two cases, the density peaks does not move significantly throughout the simulations, while the disc viscously evolves (see Fig.~\ref{fig:sigma_evolution} in the Appendix for the complete time evolution of the density profiles). Fig.~\ref{fig:warp} shows the radial tilt profiles for all 12 simulations at 1000 binary orbits. In most cases, the disc warps by only a few degrees and is otherwise smooth. The exception to that is the disc $i=60^\circ$ inclination and low mass ($M_{\rm d}=0.001$) which shows a break, as seen also in Fig.~\ref{fig:sigma} and Fig.~\ref{fig:splash} 

The snapshots shown in Fig. \ref{fig:splash} also show that in the case where the disc breaks, the inner disc shows a very clear two-armed open spiral, as opposed to the tightly wound structure observed for more modestly inclined discs. This may be suggestive that strong inclinations are also present in systems such as HD 135344B \citep{Garufi13,Cazzoletti16}, that shows an open spiral and hints of shadowing \citep{Stolker16}.

Fig.~\ref{fig:tilt} shows the time evolution of the inclination angle of the disc with respect to the binary at a distance of 200 AU from the binary center of mass. We see that for small initial misalignments (top two panels) the disc inclination oscillates periodically, with a period that decreases with increasing disc mass. On top of this oscillation we also notice a slow decline towards alignment, on a timescale of at least tens of thousands binary orbit, that is probably longer than the expected disc lifetime. Polar alignment is only clear in 3 cases; namely the low and medium disc mass for the $i=60^\circ$ inclination, and the low mass for the $i=45^\circ$ one. This is expected from equation~(\ref{eq:incl}) (see also Fig.~\ref{fig:critical}) and in agreement with the results of \citet{AlyEtal2015} and \citet{Martin&Lubow2017}. The medium mass and $i=45^\circ$ case is also close to polar-alignment, but its evolution is too slow to be captured in the lifetime of this simulation. We also see clear signs for disc breaking in the $i=60^\circ$ low mass case, which is reflected in its density profile in Fig.~\ref{fig:sigma}. 

The fit to \citep{Khoeler2011} orbital solutions requires an inclination of $i\sim25^\circ$. We can see from Fig.~\ref{fig:tilt} that the most preferred initial configurations are those with initial inclination of $i=30^\circ$. These cases spend the entire simulation time oscillating very close to the inclination required by the observational fit, as well as have a density peak very close to the observed dust ring. Note that the $30^{\circ}$ case reaches the dangerous critical misalignment that could lead to a polar configuration. We thus conclude that the initial misalignment in GG Tau A cannot have been much larger than $30^{\circ}$. 

We also note that we do not observe high disc eccentricities nor an offset between the disc and binary centres of mass, except for the case where the disc breaks and polar-align, which is consistent with the observations.

\begin{figure*}
  \begin{center}
    \resizebox{80.0mm}{!}{\mbox{\includegraphics[angle=-90]{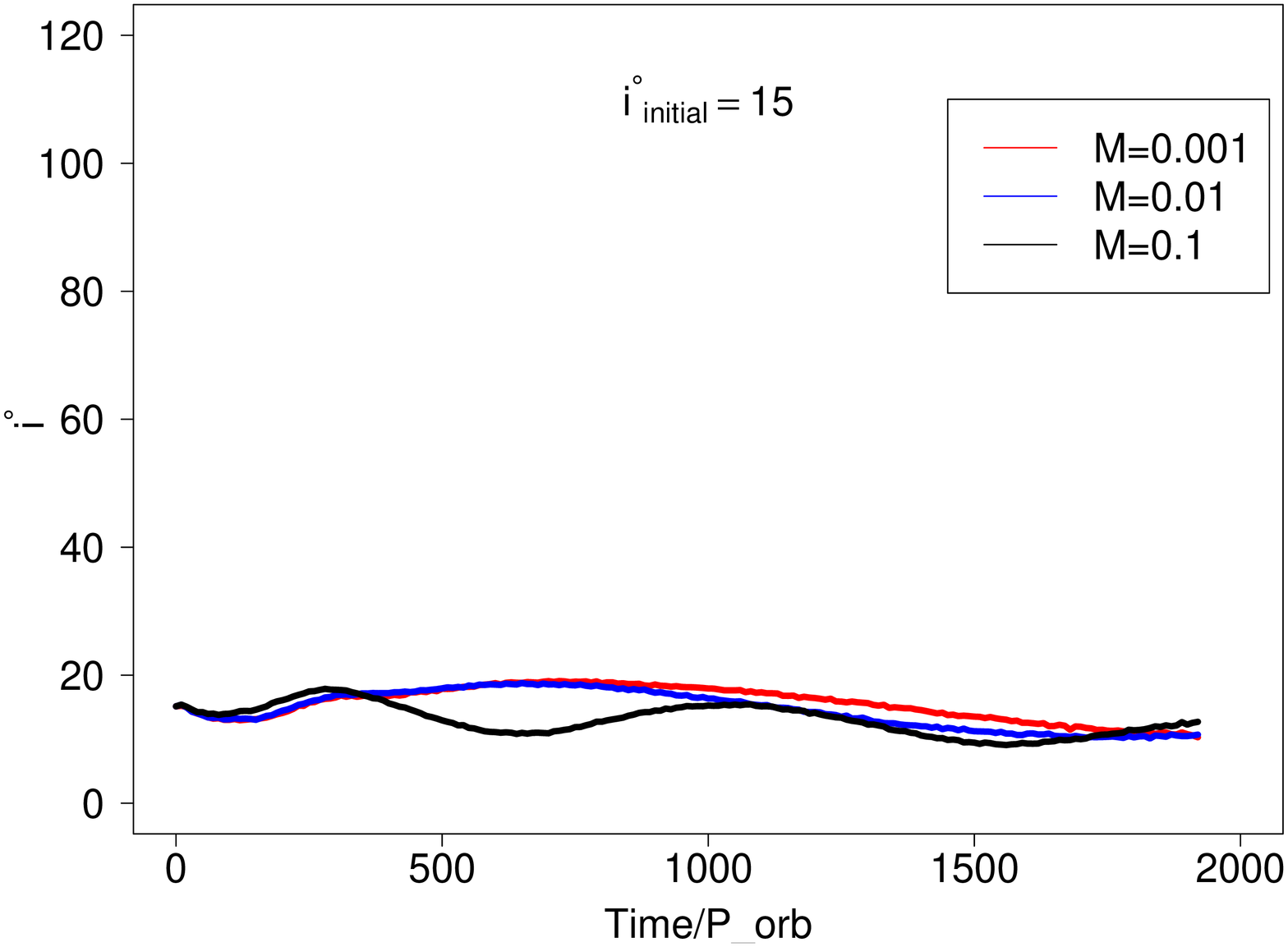}}} 
    \resizebox{80.0mm}{!}{\mbox{\includegraphics[angle=-90]{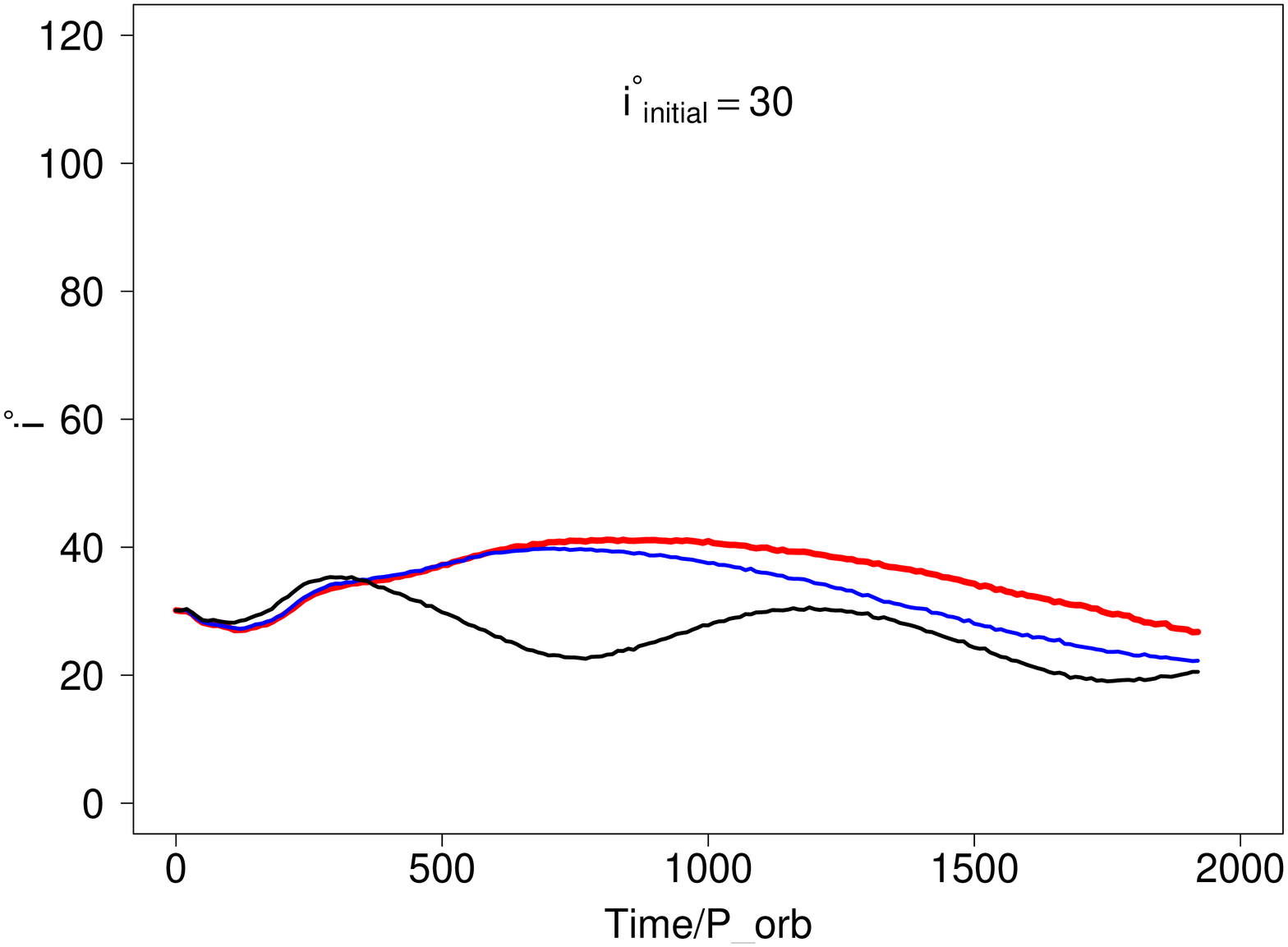}}} 
    \resizebox{80.0mm}{!}{\mbox{\includegraphics[angle=-90]{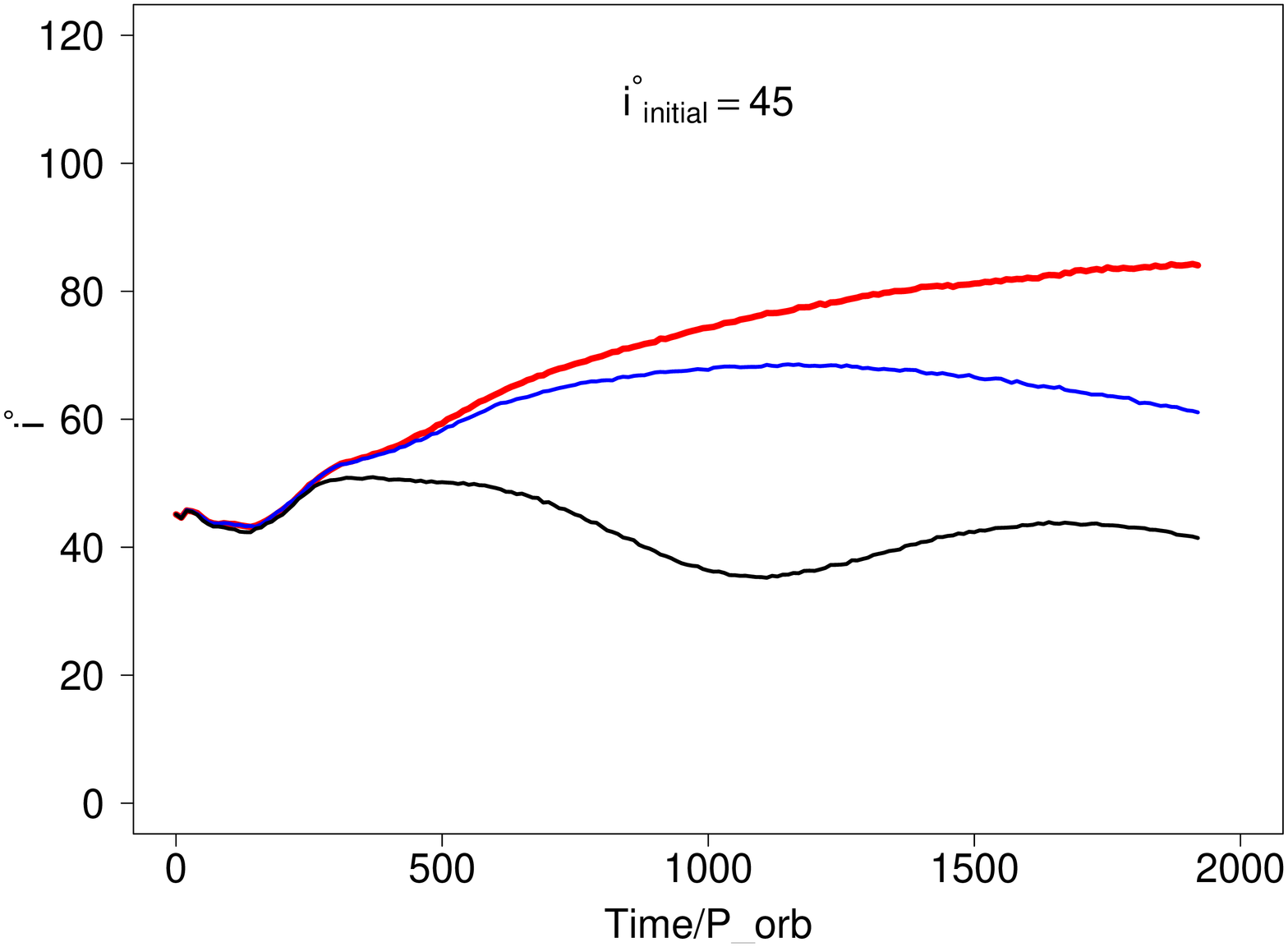}}}
    \resizebox{80.0mm}{!}{\mbox{\includegraphics[angle=-90]{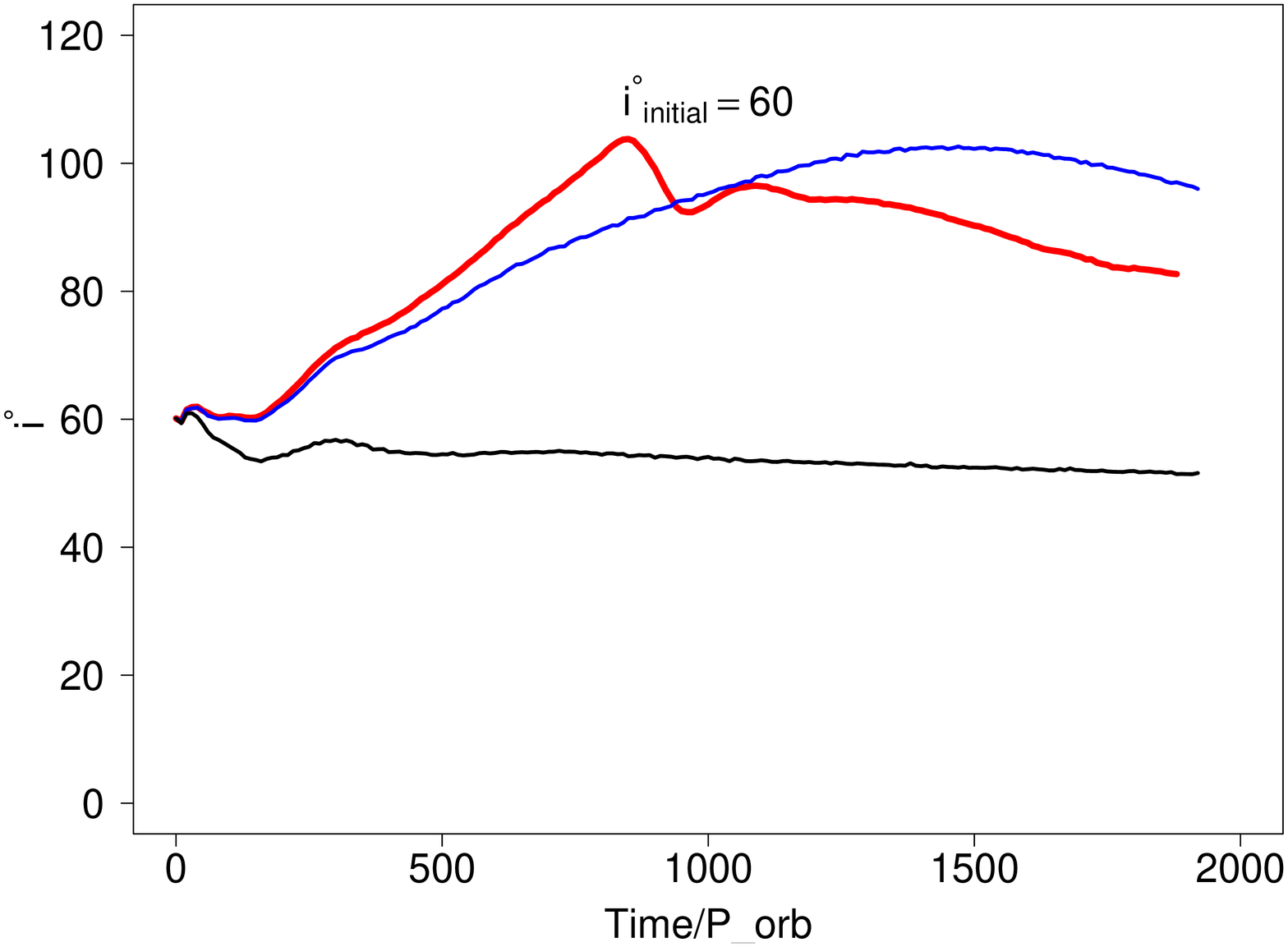}}}
 \caption{\label{fig:tilt}
    Time evolution of the inclination of a ring located at 200 AU for all 12 simulations}
  \end{center}
\end{figure*}

\begin{figure}
\resizebox{80.0mm}{!}{\mbox{\includegraphics[angle=-90]{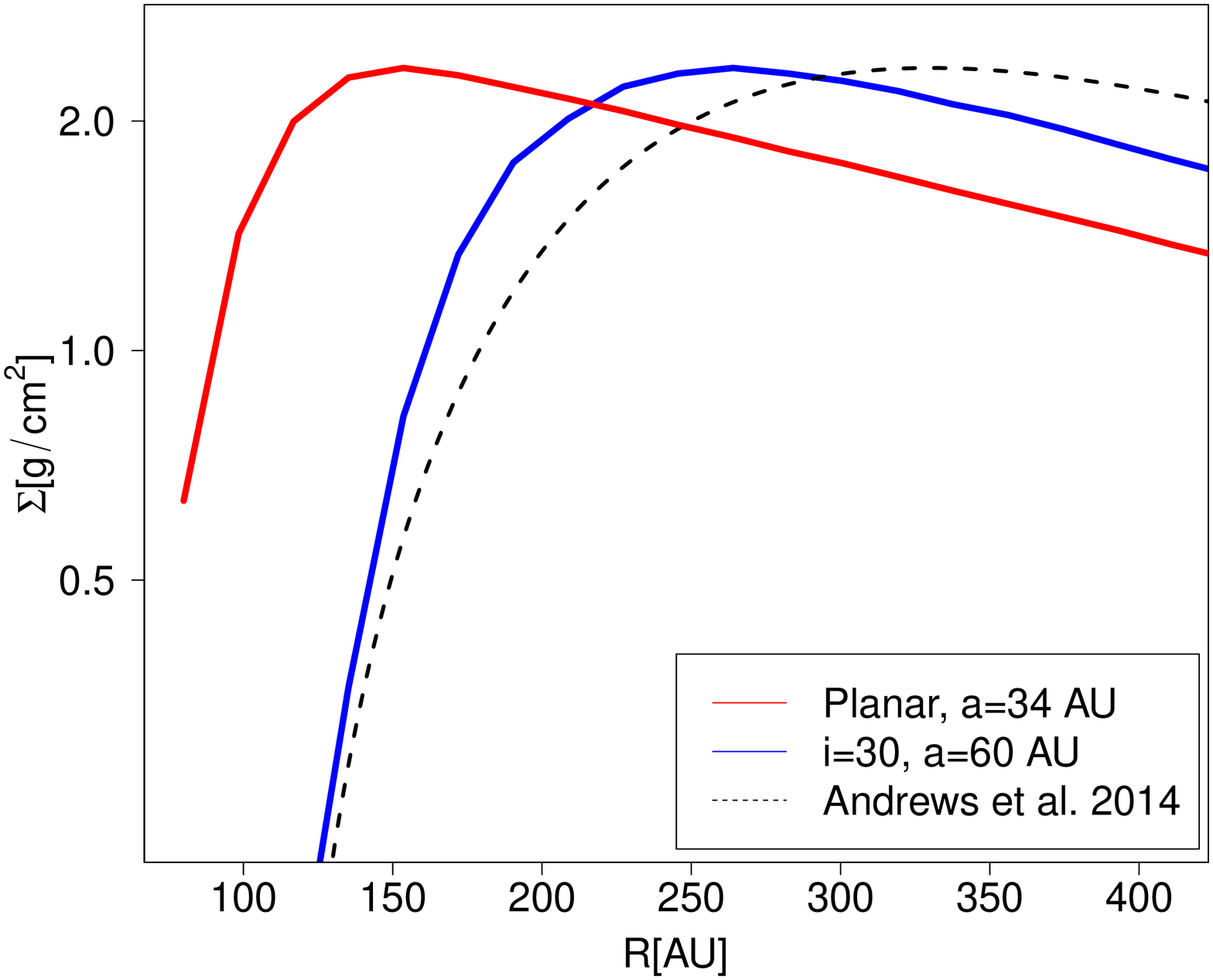}}}
\caption{\label{fig:comparison}
    Comparison between the surface density profiles of the co-planar case and $a=34$ AU, our preferred model with $i=30^\circ$, $a=60$ AU, and $M_{\rm d}=0.01$M$_\odot$, and the gas surface density profile proposed by \citet{AndrewsEtal2014}}
\end{figure}

We show in Fig.~\ref{fig:comparison} a comparison between the density profiles of a planar disc with $a=34$ AU, an inclined disc with medium mass ($M_{\rm d}=0.01$), $a=60$ AU, and $i=30^\circ$, and the gas density profile inferred from the dust observation by \citet{AndrewsEtal2014}. It is clear that an inclined disc consistent with the astrometric data fits the observations much better than a planar one.

From the observational point of view, it is worth noting that strong asymmetries are observed both in $^{13}$CO (J=3-2) emission \citep{TangEtal2016} and in near-infrared scattered light \citep{Itoh14}. Both these observables depend on the  heating and illumination from the central sources, and the asymmetric features can be ascribed to shadowing casted by dense material inside the circumbinary disk cavity. This is consistent with a complex geometry as the one we propose, with some material orbiting on a plane different than that of the circumbinary disk.

\section{Conclusion}
\label{sec:conclusion}
We presented a model for the GG Tau A system in an effort to explain the recent observations of a dust ring orbiting the binary. Previous investigations considered a co-planar gas disc and found that the tidal truncation radius is significantly smaller than the radius at which the dust ring is observed. We chose a binary eccentricity and semi major axis consistent with the observational orbital data and investigated whether this scenario would alleviate the conflict with the dust observation.

We ran a total of 12 SPH simulations for 3 different mass values each with 4 disc-binary initial misalignment angles. The surface density profiles of all cases except those for the highest inclination show a density peak very close to the radius of the observed dust ring. The reason for the discrepancy of the highly inclined case ($i=60^\circ$) is that, as expected from theoretical predictions, the binary-disc interaction causes the disc to polar-align around the binary eccentricity vector, allowing the cavity to shrink further. We expect this to happen for the $i=45^\circ$ case as well if the simulation is run long enough, as this inclination undergoes a very slow evolution due to its proximity to the transition between the polar and azimuthal precession regimes. 

The disc inclination profiles put more constraints on our parameter space. We showed that a disc with an initial inclination of ($i=30^\circ$) provides the best model for the observation as the disc oscillates for the entire simulation time (a significant fraction of the expected disc lifetime) around values close to $i\sim25^\circ$ obtained from the fit of the astrometric measurements. We showed that this configuration is stable against tearing for the disc parameters chosen in our study. We compared our preferred model to a gas model obtained by fitting the dust observations and indeed obtained good agreement.

It is worth noting that in this paper we only considered the evolution of a gas disc. Dust gas interactions, especially in the presence of binary induced precession and warping, is key to understanding the dynamics of the system and provide better constrains to our model. The dynamics of dust in a strongly warped and precessing system has never been properly considered in the literature and we plan to explore these issues in a future investigation.

\bibliographystyle{mnras} \bibliography{refs}
\appendix
\section{Evolution of Density Profiles}
\label{app:sigma_evolution}
\begin{figure}
\resizebox{80.0mm}{!}{\mbox{\includegraphics[angle=-90]{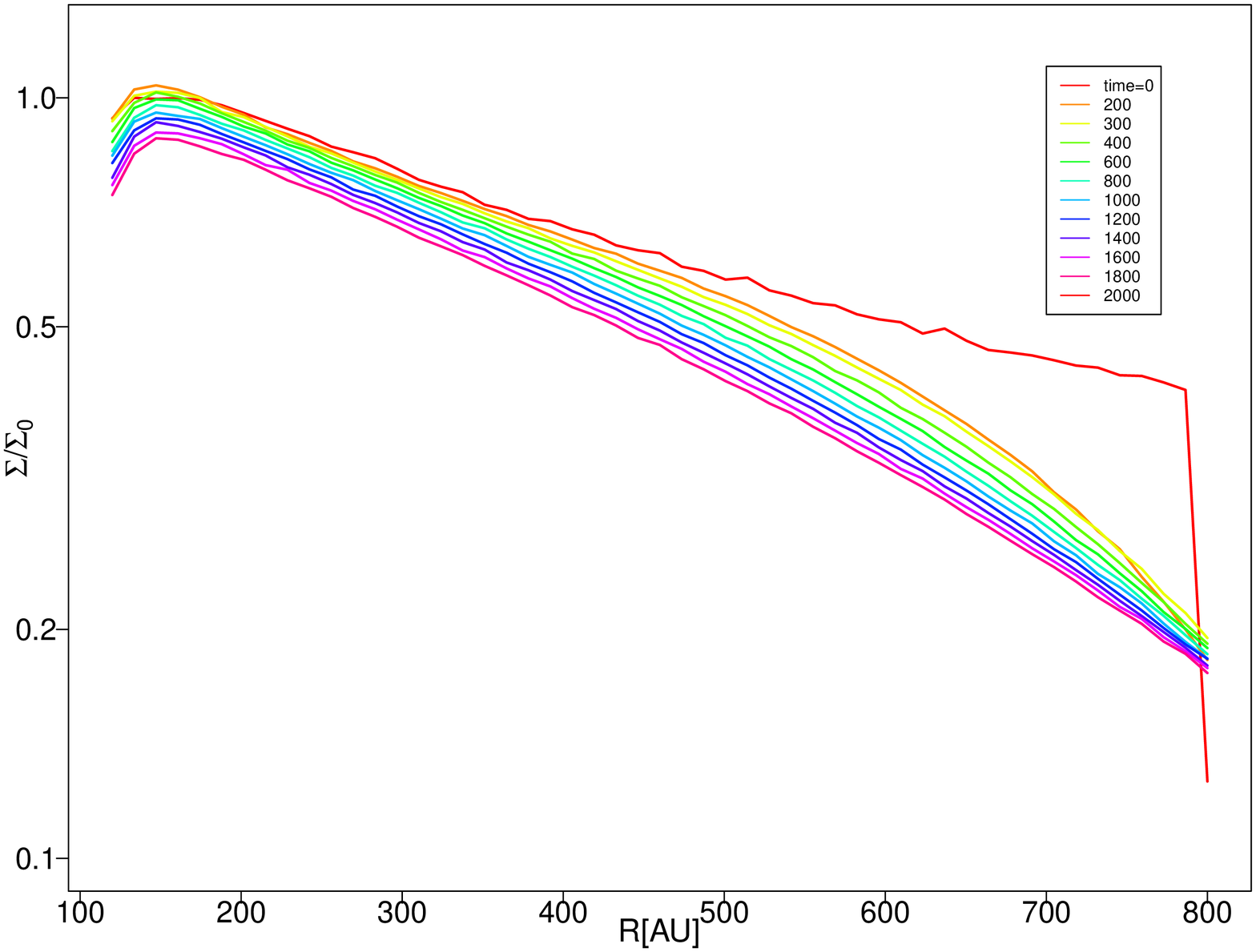}}}
\caption{\label{fig:sigma_planar}
    Surface density profiles at different times for a co-planar disc around the GG Tau A binary with $a=34$ AU and $e=0.28$}
 \end{figure}

Fig.~\ref{fig:sigma_planar} shows the density profile for a co-planar disc with a binary semi major axis and eccentricity of $a=34$ and $e=0.28$. Here the density peaks at a radius $\lesssim$ 180 AU, similar to the findings of \citet{CazzolettiEtal2017} and further confirming that a co-planar does not appear to explain the observations

\begin{figure*}
  \begin{center}
    \resizebox{40.0mm}{!}{\mbox{\includegraphics[angle=-90]{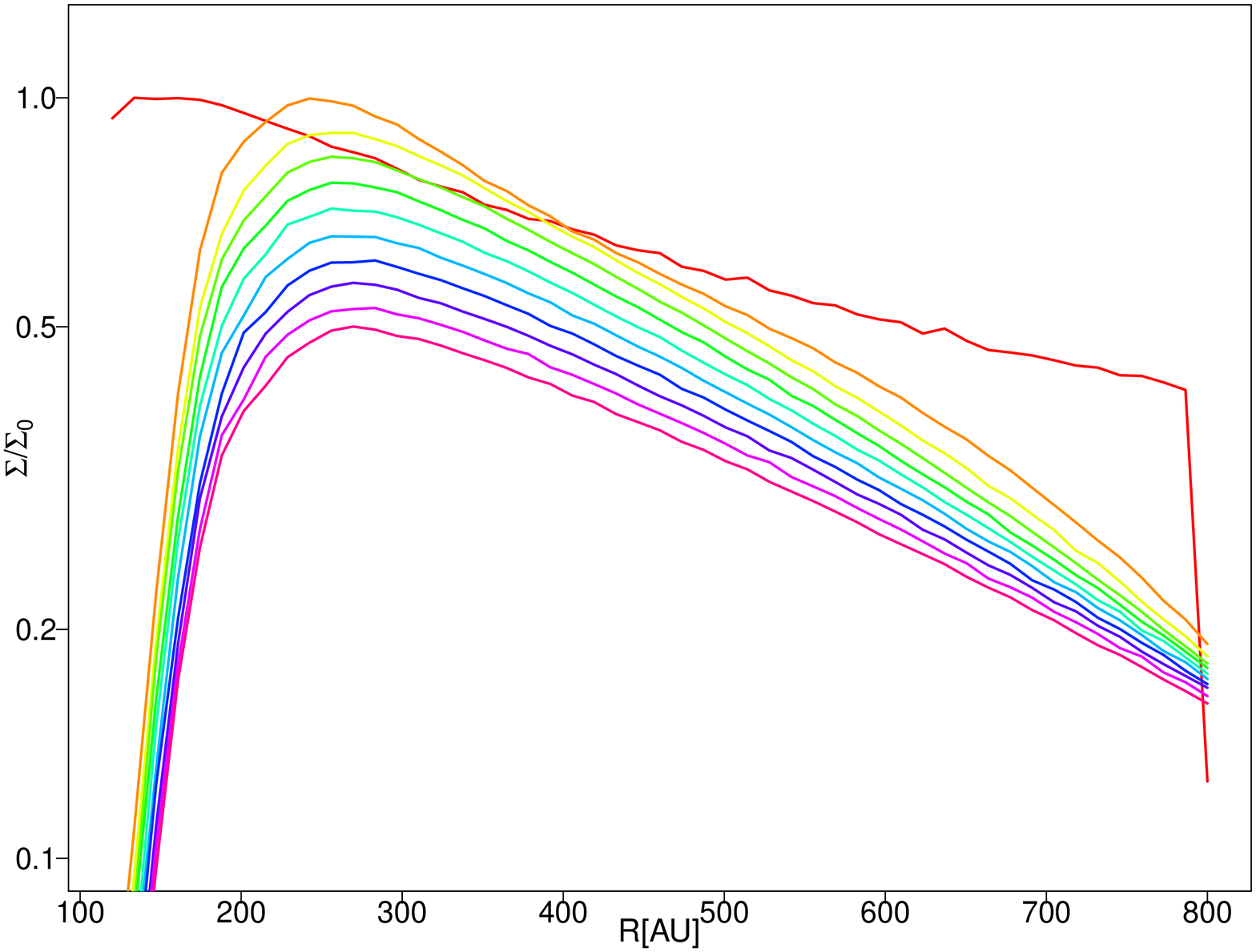}}} 
    \resizebox{40.0mm}{!}{\mbox{\includegraphics[angle=-90]{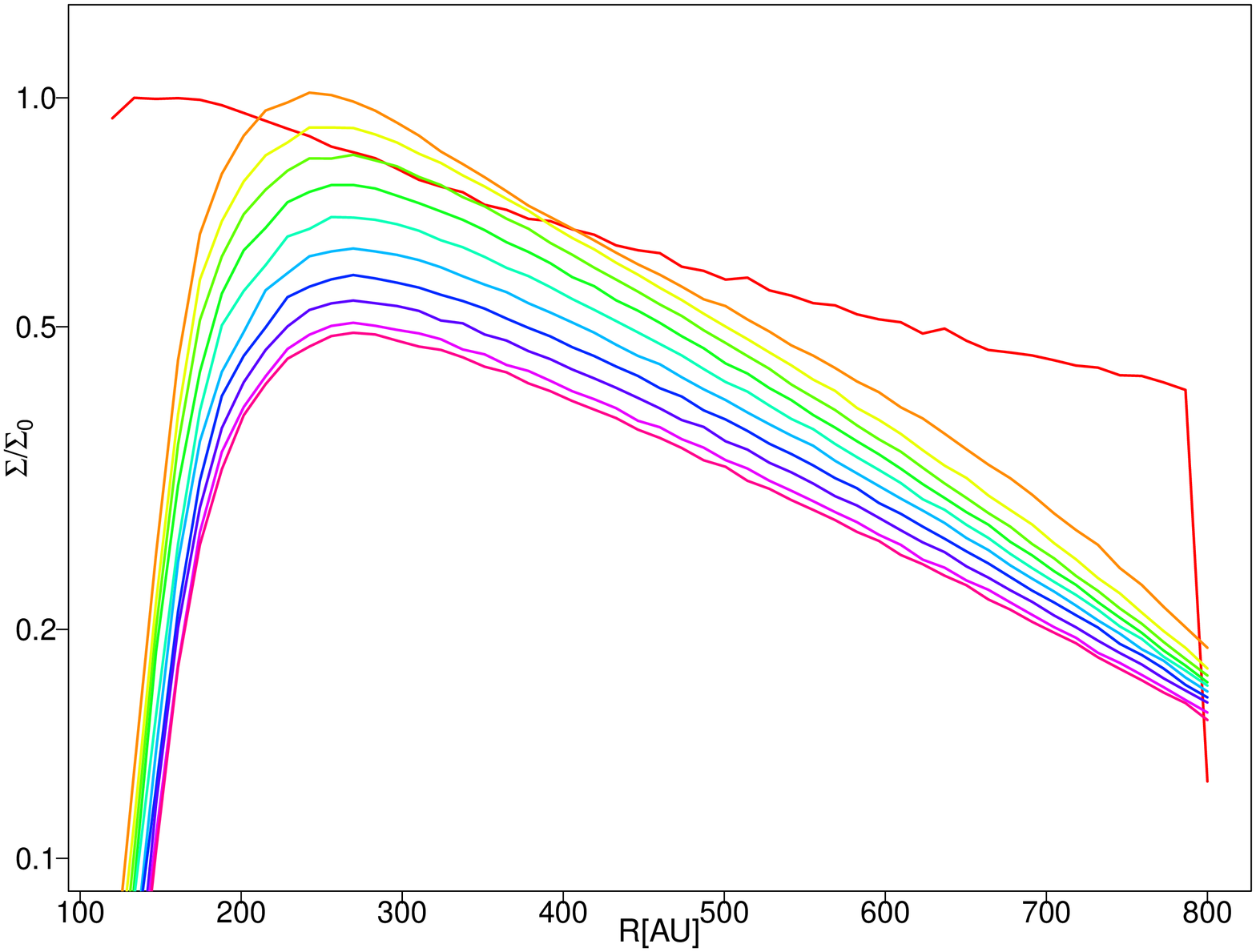}}} 
    \resizebox{40.0mm}{!}{\mbox{\includegraphics[angle=-90]{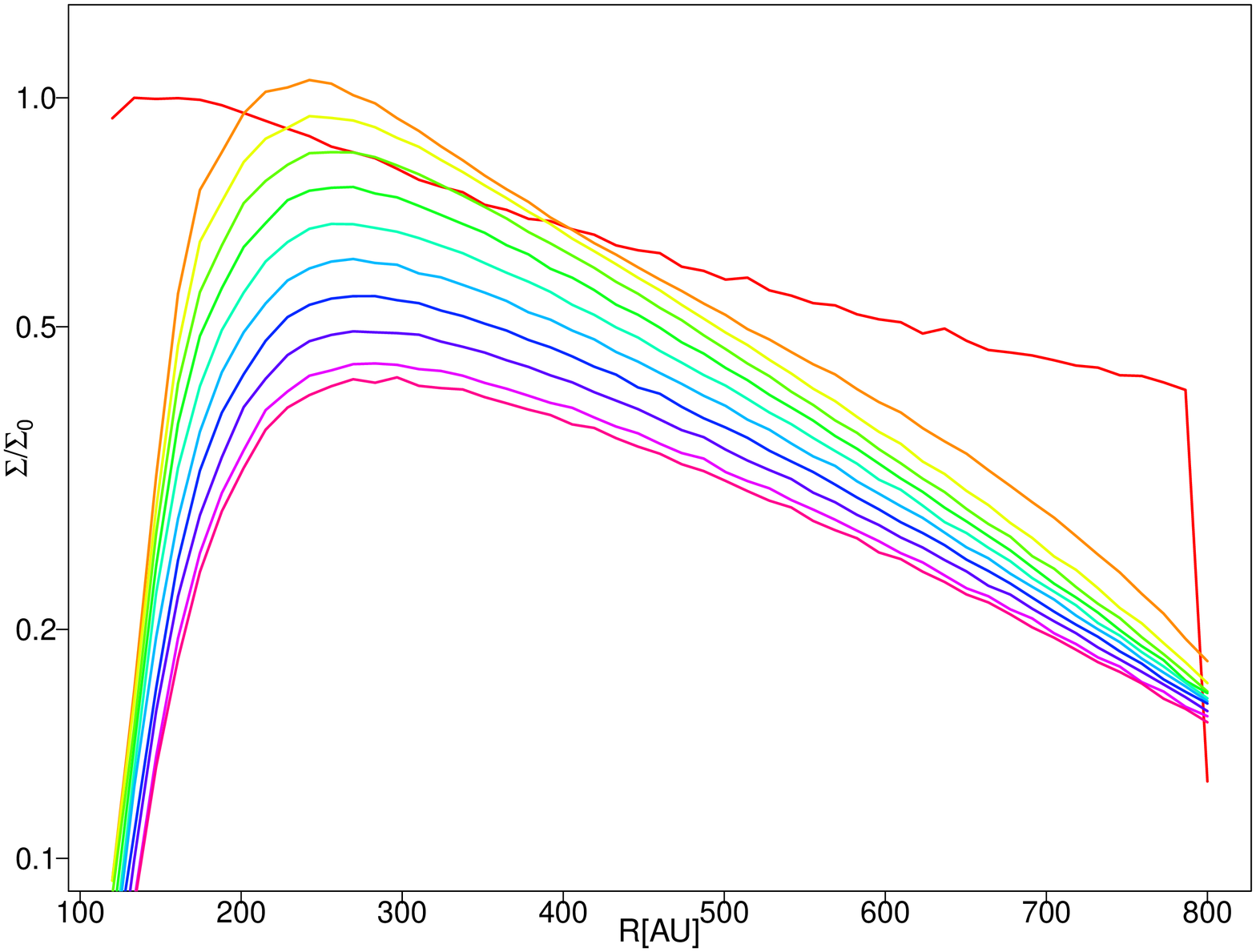}}}
    \resizebox{40.0mm}{!}{\mbox{\includegraphics[angle=-90]{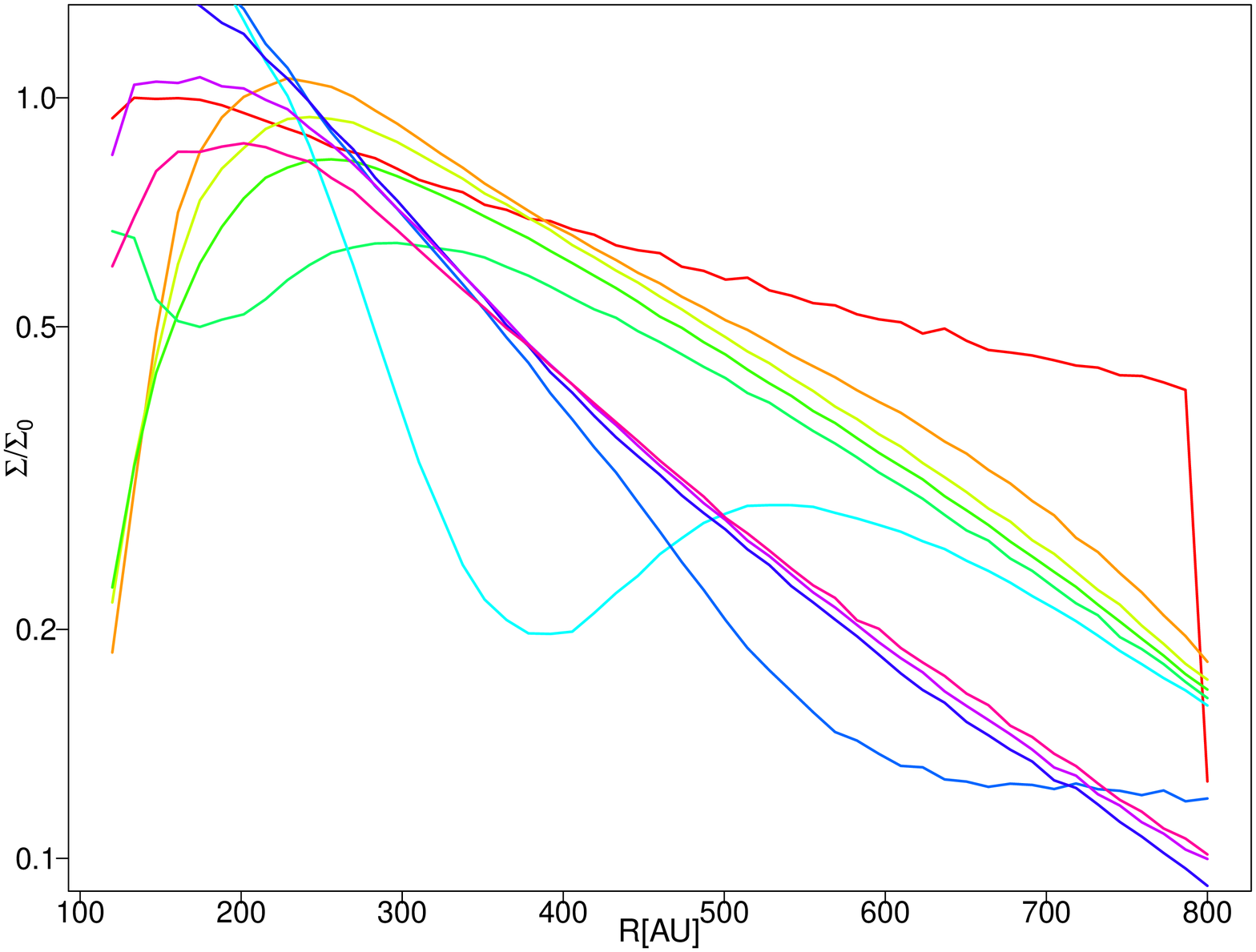}}}
    \resizebox{40.0mm}{!}{\mbox{\includegraphics[angle=-90]{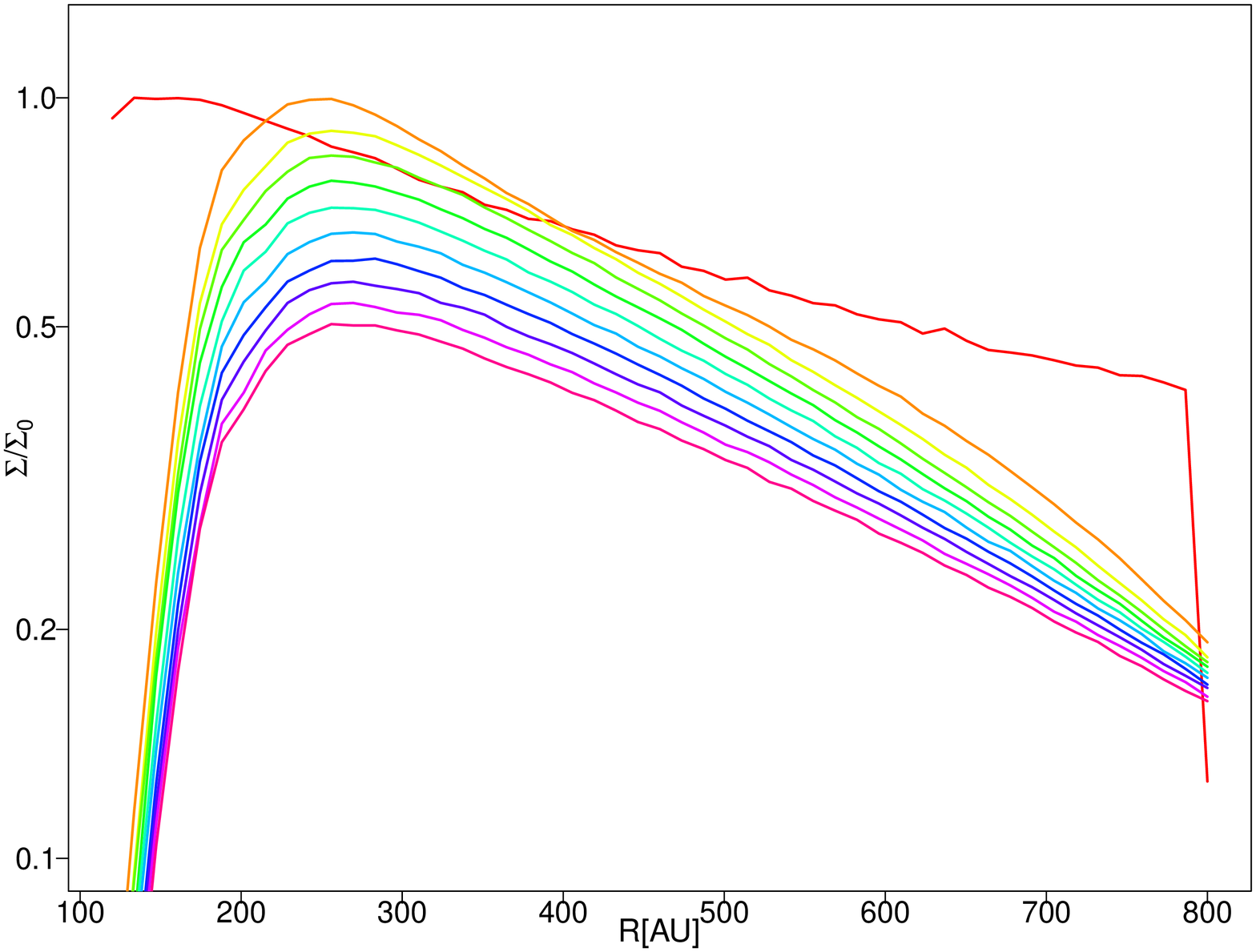}}}
    \resizebox{40.0mm}{!}{\mbox{\includegraphics[angle=-90]{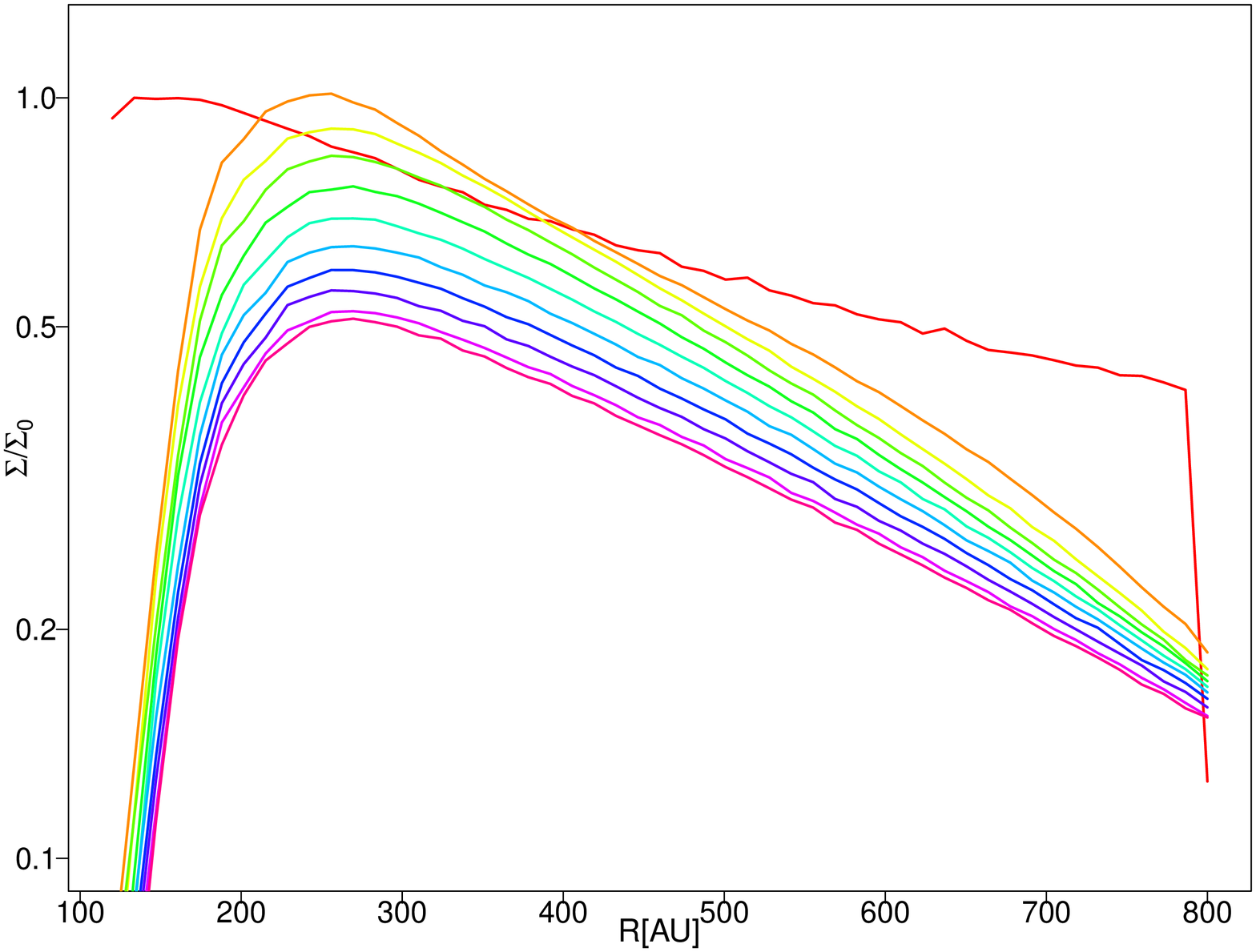}}}
    \resizebox{40.0mm}{!}{\mbox{\includegraphics[angle=-90]{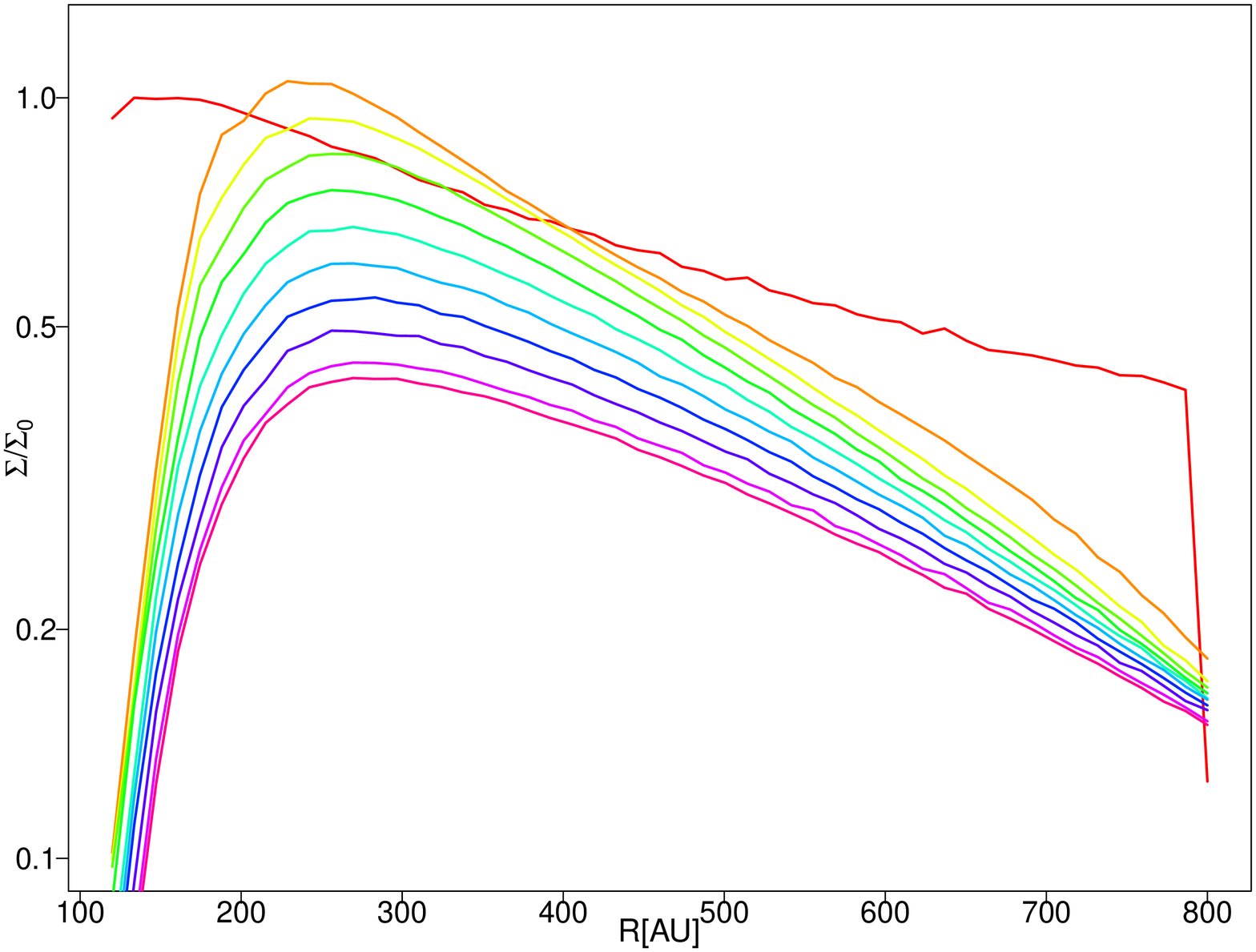}}}
    \resizebox{40.0mm}{!}{\mbox{\includegraphics[angle=-90]{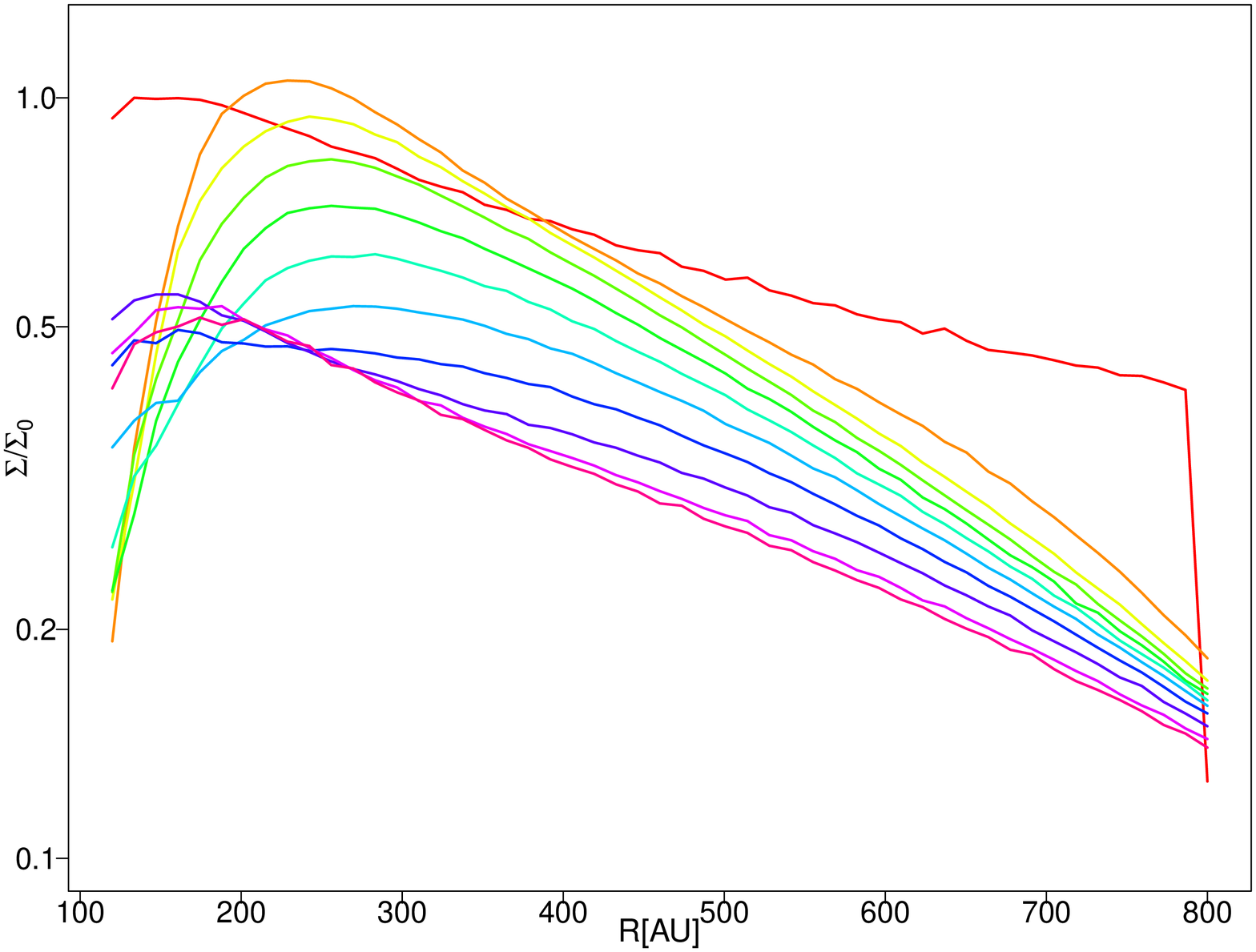}}}
    \resizebox{40.0mm}{!}{\mbox{\includegraphics[angle=-90]{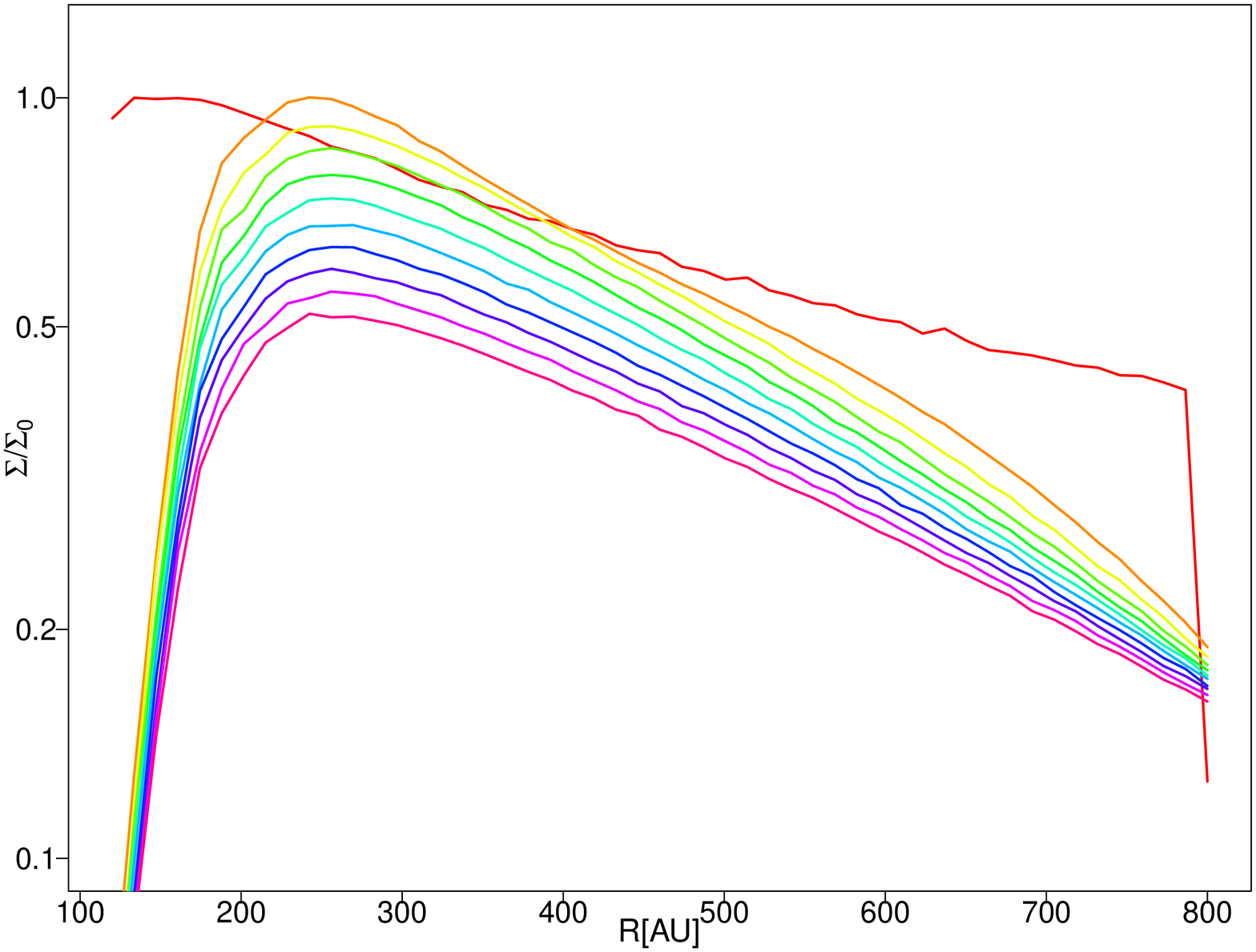}}}
    \resizebox{40.0mm}{!}{\mbox{\includegraphics[angle=-90]{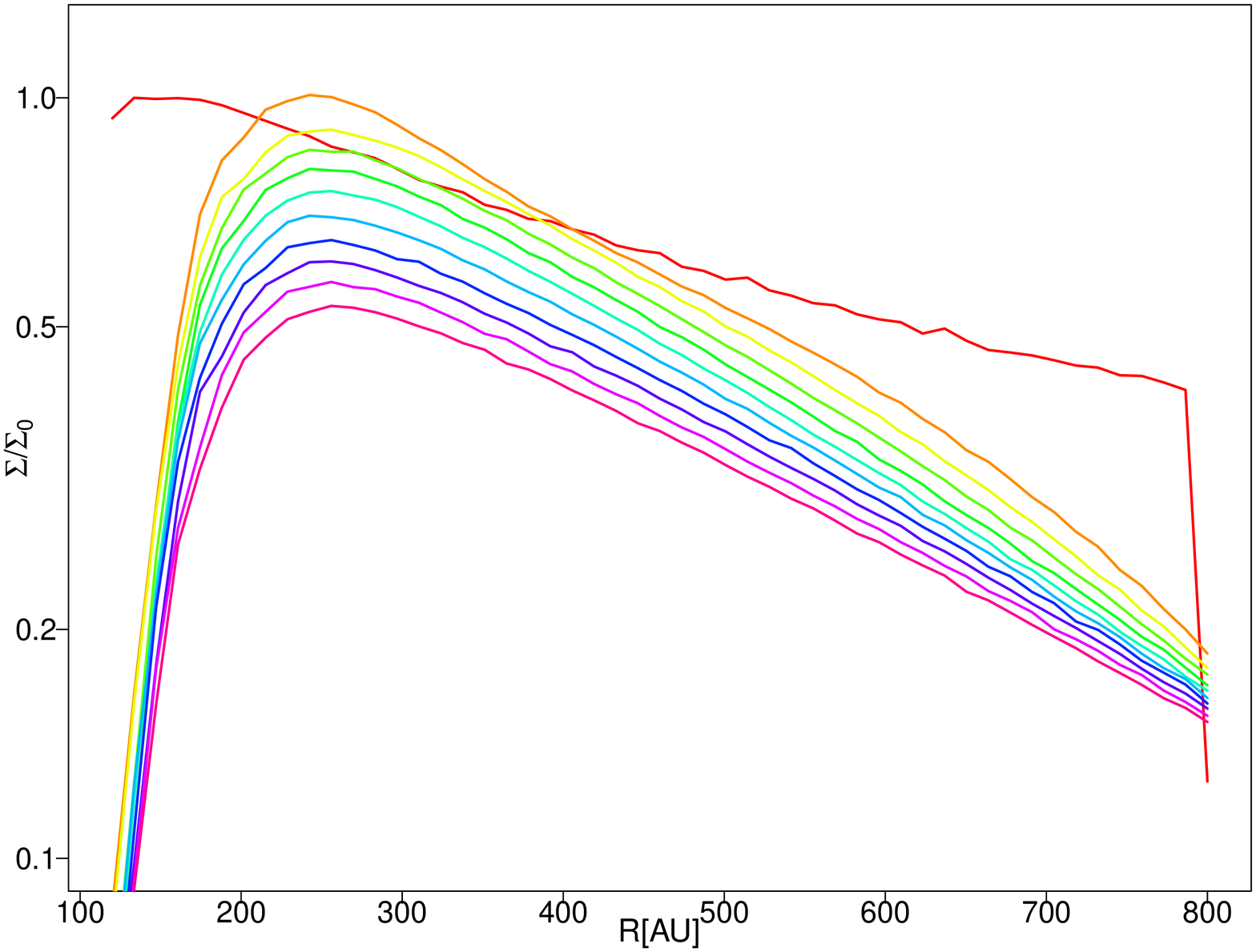}}}
    \resizebox{40.0mm}{!}{\mbox{\includegraphics[angle=-90]{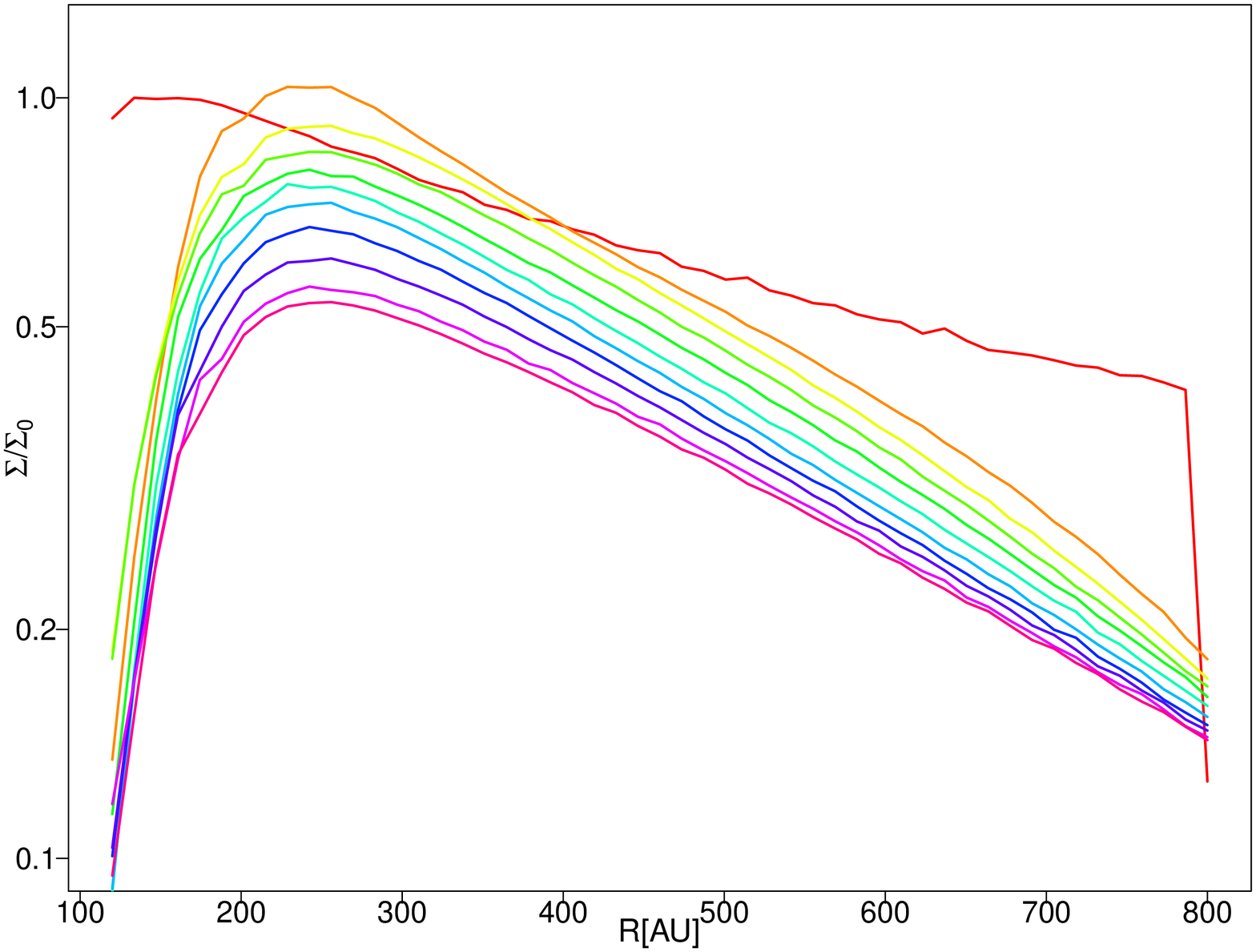}}}
    \resizebox{40.0mm}{!}{\mbox{\includegraphics[angle=-90]{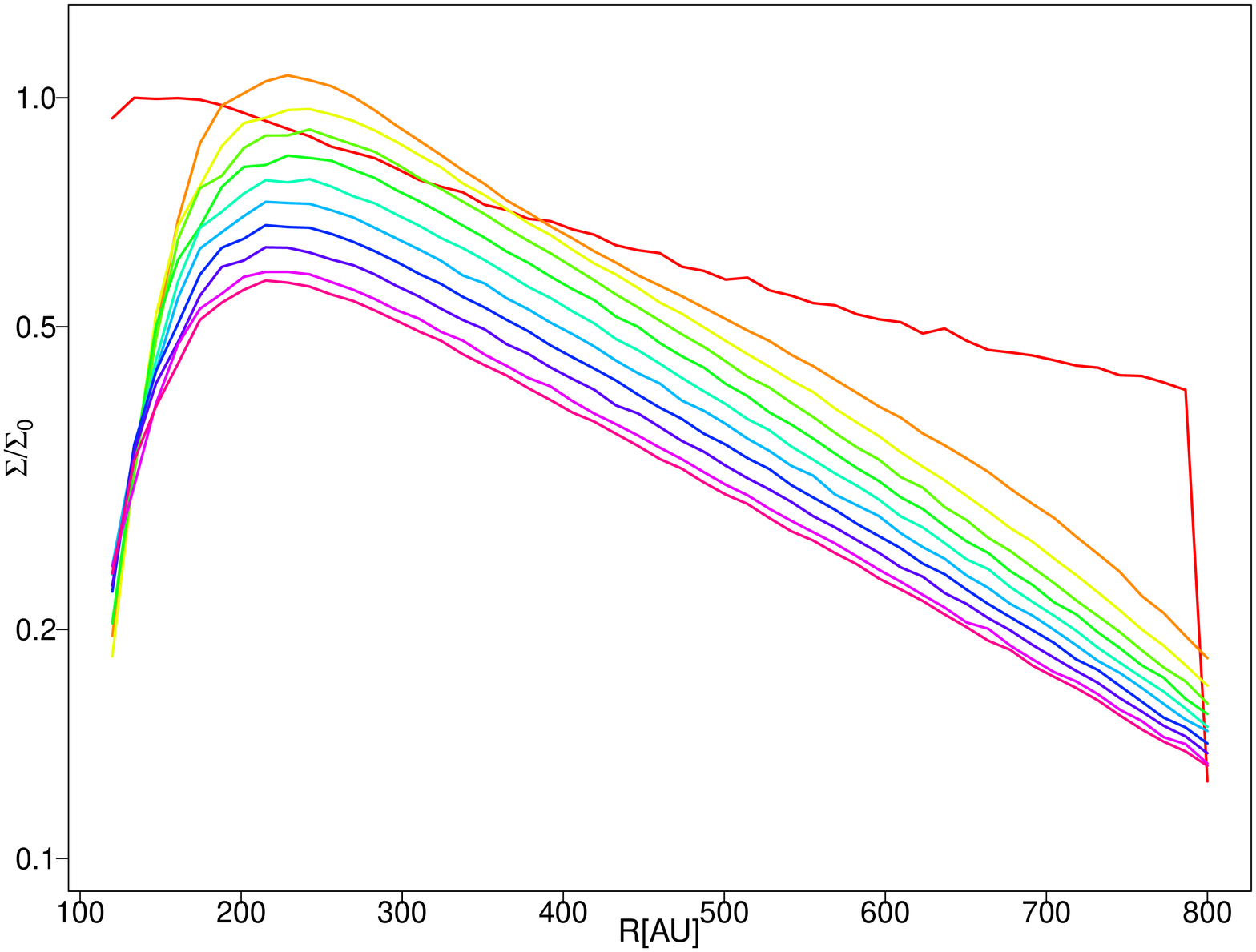}}}
 \caption{\label{fig:sigma_evolution}
    Radial surface density profiles at different times for the entire parameter space. From left to right, panels show the results for disc inclinations of $i=15^\circ$, $30^\circ$, $45^\circ$, $60^\circ$. From top to bottom, the results are shown for disc masses $M_d=0.001$, 0.01, 0.1 M$_\odot$. The binary parameters in all panels are $a=60$ AU and $e=0.45$}
  \end{center}
\end{figure*}

Fig.~\ref{fig:sigma_evolution} presents the entire time evolution of  radial density profiles for the entire simulations suite (12 in total). From left to right panels show increasing initial inclinations ($i=15^\circ$, $30^\circ$, $45^\circ$, $60^\circ$). Rows from top to bottom show different disc masses ($M_d=0.001$, 0.01, 0.1)M$_\odot$. Each panel shows the evolution of the density profile at different times from t=0 to t=2000 binary orbits (with increments of 200 binary orbits, see legend in Fig.~\ref{fig:sigma_planar}). We can see that viscous evolution of the disc does not change the location of the density peak, Which occurs at a distance consistent with the dust observations for all cases except for the disc with $i=60^\circ$ inclination and low and medium masses.

\label{lastpage}
\end{document}